%% file: main.tex
\documentclass[reprint,superscriptaddress,aps,prd,nofootinbib]{revtex4-2}

\usepackage{lineno}
\usepackage{graphicx}
\usepackage{amsmath,amssymb}
\usepackage{verbatim}
\usepackage{color}
\usepackage{booktabs}
\usepackage{hyperref}
\usepackage{fancyhdr, fancybox, empheq}
\usepackage{subfigure}
\usepackage{float}
\usepackage{url}
\usepackage[dvipsnames]{xcolor}
\usepackage{mathtools}
\usepackage[normalem]{ulem}
\usepackage{comment}
\usepackage{xspace}
\usepackage{bm}
\usepackage{multirow}
\usepackage{dcolumn}
\usepackage{orcidlink}

\makeatletter
\renewcommand{\paragraph}{\@startsection{paragraph}{4}{\z@}
  {1.5ex}
  {1.0ex}
   {\normalfont\small\itshape}}
\makeatother

\definecolor{sptcol}{HTML}{386A7A}
\hypersetup{
colorlinks=true,
  linkcolor=sptcol,
  citecolor=sptcol,
  urlcolor=sptcol
}

\newcommand{\agora}{\textsc{Agora}}
\newcommand{\camb}{\textsc{CAMB}}
\newcommand{\nhat}{\hat{n}}
\newcommand{\nside}{$N_{\rm side}$}
\newcommand{\planck}{{\it Planck}}

\newcommand{\ukam}{$\mu {\rm K}\textnormal{-}{\rm arcmin}$}

\begin{document}

\title{SPT-3G D1: Foreground-Robust Lensing Templates for Primordial Gravitational Wave Searches}

\input{authors}

\date{\today}

\begin{abstract}
Gravitational lensing of the cosmic microwave background (CMB) generates $B$-mode polarization that acts as a source of contamination to searches for $B$ modes generated by primordial gravitational waves (PGWs).
The strongest constraint on PGW $B$ modes is already significantly limited by lensing $B$ modes, as shown in the most recent BICEP result.
In this work, we present CMB lensing $B$-mode templates constructed using SPT-3G and \planck\ data, which characterize the lensing $B$ modes and can be used to improve PGW $B$-mode searches.
We use SPT-3G data from the 2019 and 2020 observing seasons for the $E$ modes and the CMB-reconstructed lensing potential,
and a cosmic infrared background (CIB) map from \planck\ as an external lensing tracer.
To test for extragalactic foreground biases in the lensing template, we consider CMB lensing reconstruction variants with different levels of foreground immunity: the standard and profile-hardened global minimum variance (GMV) quadratic estimators, and a polarization-only quadratic estimator.
We validate the template construction using Gaussian simulations and \agora\ simulations with realistic non-Gaussian foregrounds.
From simulations, we find that foreground-induced biases are strongly suppressed for the template constructed with the profile-hardened GMV + CIB tracer, with residual bias below 10\% of the statistical uncertainty on the template power spectrum.
Data difference tests on this template similarly show no evidence for significant foreground contamination.
Using simulations, we predict that this foreground-immune lensing template achieves a residual delensed $BB$ power of $A_{\rm lens}^{\rm res} \simeq 0.48$ averaged over $20 \leq \ell \leq 200$, the highest delensing efficiency of a lensing template to date.
These results demonstrate and validate a method to construct foreground-robust lensing templates which will be used in upcoming delensed PGW $B$-mode analyses of BICEP data.
\end{abstract}

\maketitle

\section{Introduction} \label{sec:introduction}

Measurements of large-scale polarization of the cosmic microwave background (CMB) provide one of the most powerful tests of inflation in cosmology. In particular, the detection of degree-scale $B$-mode polarization sourced by primordial gravitational waves (PGWs) would provide direct evidence for tensor perturbations generated in the early universe, and could point to the energy scale of inflation in simple models. 
Over the past decade, experiments including BICEP~\cite{bk18}, the South Pole Telescope (SPT)~\cite{lowell_bb}, POLARBEAR~\cite{polarbear_r}, SPIDER~\cite{spider_r}, \planck~\cite{tristram_r, planck_r}, and the Atacama $B$-mode Search~\cite{abs} have progressively improved sensitivity to this signal. The current state-of-the-art constraint on the tensor-to-scalar ratio, $r$, comes from the BICEP collaboration~\cite{bk18}.

The detection of PGWs is complicated by the presence of other sources of $B$-mode polarization, such as polarized Galactic foregrounds and gravitational lensing. Since Galactic foregrounds have different frequency dependence than the CMB, they can be mitigated through multi-frequency observations. As the sensitivity of multi-frequency CMB experiments continues to improve, contamination from gravitational lensing, if left unmitigated, becomes the dominant contributor to the total uncertainty on $r$, $\sigma(r)$. This is already the case for the most recent BICEP analysis, where lensing sample variance dominates $\sigma(r)$~\cite{bk18}.

Gravitational lensing of the CMB by large-scale structure changes the polarization patterns of the CMB, mixing $E$ and $B$ modes and thereby turning some $E$ modes into $B$ modes. This produces a $B$-mode signal that is indistinguishable from primordial $B$ modes in frequency space. Addressing the impact of this effect on $r$ requires estimating and accounting for the lensing-induced $B$-mode pattern, which is a process commonly referred to as ``delensing."
The reduction of $\sigma(r)$ with delensing was first demonstrated on data by the BICEP and SPT collaborations in~\cite{bkspt_delens}.

Several approaches to constructing $B$-mode lensing templates have been explored in recent years. One strategy is to use an estimate of the lensing potential $\phi$ to remap the observed polarization fields~\cite{planck_lensing_2018, polarbear_delensing_2019, bkspt_delens}.
Another strategy, which we adopt, is to construct the template by combining an estimate of the lensing potential with measurements of the CMB $E$-mode polarization, which at gradient order takes the form of a harmonic-space convolution~\cite{quadratic_delens_2012, manzotti_2017, anton_quadratic_delens_2021, act_lensing_template}.
Once constructed, the lensing template can be used either to subtract the lensing power in the observed polarization maps directly, or to reduce lensing sample variance through cross-correlation within a likelihood framework.

In this work, we present CMB lensing $B$-mode templates constructed with data from SPT-3G, the third-generation receiver on the SPT.
Relative to previous lensing templates constructed in \cite{bkspt_delens}, which used only the cosmic infrared background (CIB) as a $\phi$ tracer, this work incorporates substantially improved estimates of the lensing potential. In particular, we incorporate reconstructions of the lensing potential from the SPT-3G D1 dataset~\cite[][hereafter O26]{spt_lensing_20192020}, which achieves the highest per-mode signal-to-noise on the CMB lensing potential to date.
Furthermore, we combine the CMB-reconstructed lensing potential with a CIB map from \planck\, yielding a lensing template with the highest delensing efficiency to date.
The method presented and the resulting lensing templates will be included as part of the next joint BICEP and SPT $r$ analysis.

The paper is organized as follows. We begin in Sec.~\ref{sec:methods} by describing the formalism and methods used to construct the lensing $B$-mode templates. Next we outline the inputs to our analysis in Sec.~\ref{sec:inputs}. Then in Sec.~\ref{sec:results}, we present validation tests using simulations and results from data, including assessments of foreground-induced biases and delensing efficiency. We conclude in Section~\ref{sec:conclusion} with a summary and an outlook toward applications in upcoming CMB analyses.
We take the \planck\ 2018 {\tt TTTEEE\_lowl\_lowE\_lensing} cosmology~\cite{planck2018cosmo} as the fiducial cosmology in this work.

\section{Methods} \label{sec:methods}

\subsection{Lensing Template Formalism} \label{sec:formalism}

The observed CMB polarization fields are distorted by gravitational lensing from the intervening large-scale structure. To construct a template for the lensing-induced $B$ modes, we begin by formalizing how lensing modifies the CMB polarization fields and how an estimate of the projected lensing field can be used to predict the resulting $B$-mode pattern.

Gravitational lensing remaps the unlensed CMB temperature and polarization fields according to the gradient of the lensing potential $\phi$, which arises from a line-of-sight integral of the gravitational potential \cite{lewis_challinor_2006}.
For the unlensed spin-2 polarization fields $X_\pm(\hat{\bm{n}}) = Q \pm iU$, this remapping is expressed as
\begin{equation} \label{eq:remap}
    \tilde{X}_\pm(\hat{\bm{n}}) = X_\pm(\hat{\bm{n}} + \nabla\phi(\hat{\bm{n}})),
\end{equation}
where $\tilde{X}_\pm$ denotes the lensed fields.

To leading order in the lensing potential, the lensing-induced $B$ modes can be expressed in harmonic space as a convolution between the $E$-mode polarization and the lensing potential.
Expanding Eq.~\ref{eq:remap} to first order in $\phi$ (gradient-order approximation) and decomposing the polarization into $E$ and $B$ modes in spherical harmonic space yields
\begin{equation} \label{eq:B_lens}
    B^{\rm lens}_{\ell m} = \sum_{\ell' m'} \sum_{L M} (-1)^M
    \begin{pmatrix}
    \ell & \ell' & L \\
    m & m' & -M
    \end{pmatrix}
    g^{EB}_{\ell \ell' L} E_{\ell' m'} \phi_{L M}.
\end{equation}
Here, $E_{\ell' m'}$ are the $E$-mode spherical harmonic coefficients, $\phi_{LM}$ are the coefficients of the lensing potential, the Wigner-3j symbol encodes angular momentum conservation in the coupling of spherical harmonic modes, and
$g^{EB}_{\ell \ell' L}$ is the $E\phi \to B$-mode coupling kernel
(the functional form of $g^{EB}_{\ell \ell' L}$ can be found in \cite{quadratic_delens_2012}). This equation highlights that lensing transfers power from $E$ modes into $B$ modes through mode coupling with the lensing potential.

Equation~\ref{eq:B_lens} provides the basis for constructing a template of the lensing-induced $B$-mode signal.
In practice, the true lensing potential is not known and must be replaced by a tracer $\hat{\phi}$, with an appropriate Wiener filter applied to minimize the delensed $B$-mode variance~\cite{quadratic_delens_2012}. We denote the resulting Wiener-filtered tracer as $\phi^{\rm WF}$.
Also note that Eq.~\ref{eq:B_lens} is written in terms of the unlensed $E$-mode field, but the unlensed polarization pattern is not directly observable.
Instead, we construct the template using the lensed and noisy measured $E$ modes. As shown in \cite{anton_quadratic_delens_2021}, using lensed rather than unlensed $E$ modes in the gradient-order template is in fact preferable, as it enables cancellations between higher-order lensing terms that would otherwise introduce a residual delensing floor.
For the same reasons as for $\phi$, we Wiener filter the measured $E$ modes, denoted $E^{\rm WF}$.
The corresponding lensing $B$-mode template is therefore
\begin{equation} \label{eq:B_template}
    B^{\rm LT}_{\ell m} = \sum_{\ell' m'} \sum_{L M} (-1)^M
    \begin{pmatrix}
    \ell & \ell' & L \\
    m & m' & -M
    \end{pmatrix}
    g^{EB}_{\ell \ell' L} E^{\rm WF}_{\ell' m'} \phi^{\rm WF}_{LM}.
\end{equation}

The performance of this template depends on both the quality of the measured $E$-mode field and the correlation between the tracer $\hat{\phi}$ and the true lensing potential.
To quantify the latter, we use
\begin{equation} \label{eq:rho_phi}
    \rho_L^{\phi} = \frac{C_L^{\hat{\phi}\phi^{\rm in}}}{\sqrt{C_L^{\hat{\phi}\hat{\phi}} C_L^{\phi^{\rm in}\phi^{\rm in}}}},
\end{equation}
where $\phi^{\rm in}$ denotes the input (true) lensing potential in simulations. This correlation determines how well the tracer captures the true lensing modes that source the $B$-mode signal.

The resulting lensing $B$-mode template inherits this correlation: a higher $\rho_L^{\phi}$ leads to a stronger correlation between the template and the true lensing $B$ modes. We compute the correlation coefficient between the constructed lensing $B$-mode template and the input lensing $B$-mode field in simulations:
\begin{equation} \label{eq:rho_B}
    \rho_\ell^{B} = \frac{C_\ell^{B^{\rm LT} B^{\rm in}}}{\sqrt{C_\ell^{B^{\rm LT} B^{\rm LT}} C_\ell^{B^{\rm in} B^{\rm in}}}},
\end{equation}
where $B^{\rm in}$ denotes the input lensing $B$-mode field. To connect $\rho_\ell^B$ to a metric for the delensing efficiency, we form the delensed $B$-mode field by subtracting the template from the input:
\begin{equation}
    B^{\rm delens}_{\ell m} = B^{\rm in}_{\ell m} - B^{\rm LT}_{\ell m},
\end{equation}
whose power spectrum is
\begin{equation}
    C_\ell^{B^{\rm delens}B^{\rm delens}} = C_\ell^{B^{\rm in}B^{\rm in}} - 2C_\ell^{B^{\rm in}B^{\rm LT}} + C_\ell^{B^{\rm LT}B^{\rm LT}}.
\end{equation}
The residual lensing amplitude is then defined as the fraction of lensing power remaining after delensing:
\begin{equation}
    A_{\rm lens}^{\rm res}(\ell) \equiv \frac{C_\ell^{B^{\rm delens}B^{\rm delens}}}{C_\ell^{B^{\rm in}B^{\rm in}}} = 1 - 2\frac{C_\ell^{B^{\rm in}B^{\rm LT}}}{C_\ell^{B^{\rm in}B^{\rm in}}} + \frac{C_\ell^{B^{\rm LT}B^{\rm LT}}}{C_\ell^{B^{\rm in}B^{\rm in}}}.
\end{equation}
When the filters are optimal, the template auto- and cross-spectra agree, i.e., $C_\ell^{B^{\rm LT}B^{\rm LT}} = C_\ell^{B^{\rm LT}B^{\rm in}}$. Intuitively, optimal Wiener-filtering scales each mode by its signal-to-noise so that the template auto-power is brought down to exactly its cross-power with the input (as shown in Appendix~\ref{sec:filtering_and_weighting}). The residual lensing amplitude then reduces to
\begin{equation}
    A_{\rm lens}^{\rm res}(\ell) = 1 - (\rho_\ell^{B})^2.
\end{equation}
We average over a multipole range relevant for degree-scale $B$-mode measurements, $20 \leq \ell \leq 200$, to obtain
\begin{equation} \label{eq:a_lens_res}
    A_{\rm lens}^{\rm res} = \left\langle A_{\rm lens}^{\rm res}(\ell) \right\rangle_{\ell=[20,200]}.
\end{equation}
Throughout this work, we use the residual lensing amplitude defined in Eq.~\ref{eq:a_lens_res} as the metric to evaluate the delensing efficiency of our lensing templates, where smaller values correspond to more effective delensing.

\subsection{Wiener Filtering and Optimal Combination of the Lensing Tracers} \label{sec:multitracer}

To construct the $\phi$ input to the lensing template in Eq.~\ref{eq:B_template}, we need an estimate of the lensing potential $\phi$. 
In this work we Wiener filter and combine the two lensing tracers:
\begin{enumerate}
    \item the CMB quadratic estimator (QE) reconstructed $\kappa$,\footnote{The lensing convergence, $\kappa$, is related to the lensing potential, $\phi$, as $\kappa_L = \frac{L(L+1)}{2}\phi_L$. In this paper, we use them interchangeably depending on which object we use at different steps.} which we denote $I_{\rm \hat{\kappa}}$, and
    \item the CIB-based tracer of the lensing convergence, denoted $I_{\rm CIB}$.
\end{enumerate}
We follow the formalism in~\cite{yu_2017, agora} to construct a combined tracer
\begin{equation} \label{eq:combined_tracer}
    I = \sum_i c_i I_i,
\end{equation}
where the coefficients $c_i$ are scale dependent and chosen to maximize the correlation between $I$ and the true lensing convergence $\kappa$.

Defining the auto- and cross-spectra $C_{L}^{I_{i}I_{j}}$ and $C_{L}^{I_{i}\kappa}$, with $i,j \in \{\hat{\kappa}, {\rm CIB}\}$, we have the matrix of correlation coefficients
\begin{equation} \label{eq:rho_ij}
    \rho_{ij} = \frac{C_L^{I_{i}I_{j}}}{\sqrt{C_L^{I_{i}I_{i}}C_L^{I_{j}I_{j}}}}
\end{equation}
and the cross-correlation coefficients of $\kappa$ and $I_i$,
\begin{equation} \label{eq:rho_ik}
    \rho_{i\kappa} = \frac{C_L^{I_{i}\kappa}}{\sqrt{C_L^{I_{i}I_{i}}C_L^{\kappa\kappa}}}.
\end{equation}
Then the optimal weights that maximize the correlation of $I$ with $\kappa$ are given by
\begin{equation} \label{eq:combined_tracer_weights}
    c_i = \sum_j (\boldsymbol{\rho}^{-1})_{ij} \rho_{j\kappa} \sqrt{\frac{C_L^{\kappa\kappa}}{C_L^{I_i I_i}}}.
\end{equation}
Here, $\boldsymbol{\rho}^{-1}$ provides the normalized inverse-covariance weights, $\rho_{j\kappa}$ contributes to part of the Wiener filter, and the factor involving the square root converts each tracer into units of convergence. 
This combined tracer provides a better estimate of the true lensing field than either tracer alone, and yields improved delensing efficiency in the resulting lensing template.

Note that in single-tracer cases,
Eq.~\ref{eq:combined_tracer_weights} reduces to the Wiener filter of the tracer to $\kappa$.  
For the reconstructed lensing maps, it is:
\begin{equation} \label{eq:qe_wiener_filter}
    W_L^{\hat{\kappa}} = \frac{C_L^{\kappa\kappa}}{C_L^{\kappa\kappa} + N_{L}^{\kappa\kappa, (0)}},
\end{equation}
where in the denominator we have the $N_L^{(0)}$ reconstruction noise computed in O26 for each lensing estimator.
We use this filter for the QE $\hat{\kappa}$ when not combined with the CIB $\phi$ tracer.

\section{Inputs to the Lensing Template} \label{sec:inputs}

In this section we describe the inputs used to construct the lensing templates: the CMB-reconstructed lensing potential $\hat{\phi}$, the \planck\ CIB map, and the SPT-3G $E$~modes. For each, we describe the relevant formalism, data products, and simulations.

\begin{figure}[t]
	\centering
	\includegraphics[width=\columnwidth]{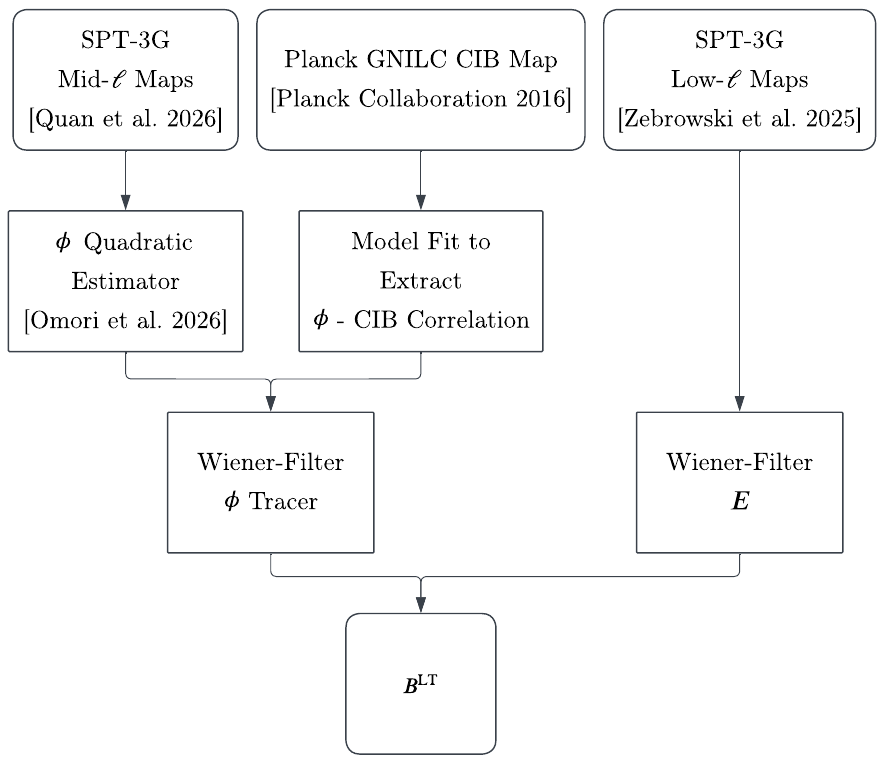}
	\centering \caption{Schematic of the lensing $B$-mode template pipeline. There are three inputs to the template $B^{\mathrm{LT}}$: the SPT-3G mid-$\ell$ maps (95/150/220~GHz), which provide the QE $\phi$; the \planck\ 545~GHz GNILC CIB map~\cite{gnilc}; and the SPT-3G low-$\ell$ maps (95/150~GHz), which provide the $E$ modes. 
    The mid-$\ell$ and low-$\ell$ map processing is described in \cite{wei_maps} and \cite{lowell_bb}, respectively; the QE reconstruction is described in O26~\cite{spt_lensing_20192020}.
    The QE reconstruction and CIB tracers are combined and Wiener filtered into a $\phi$ tracer (Sec.~\ref{sec:multitracer}). The low-$\ell$ maps are calibrated (Sec.~\ref{sec:E_mode_data}) and $\mathbb{C}^{-1}$ filtered (Sec.~\ref{sec:E_mode_filtering}) to form the Wiener-filtered $E$ modes. The lensing template $B^{\rm LT}$ construction from the Wiener-filtered $E$ and $\phi$ are given by Eq.~\ref{eq:B_template}.
    }
	\label{fig:pipeline_diagram}
\end{figure}

For the QE reconstructed $\phi$ and $E$ modes, we use SPT-3G maps built from observations of the Main 1500~deg$^2$ field during the 2019 and 2020 observing seasons. 
The SPT-3G Main field spans right ascensions~(RA) between $20^\textrm{h}40^\textrm{m}0^\textrm{s}$ and $3^\textrm{h}20^\textrm{m}0^\textrm{s}$ and declinations~(dec) from $-42^\circ$ to $-70^\circ$.
These data are processed into two distinct map sets optimized for different angular scales: a mid-$\ell$ map set \cite{wei_maps}, which serves as input to the QE $\phi$ reconstruction, and a low-$\ell$ map set \cite{lowell_bb}, from which the $E$ modes for the lensing template are derived.
For the external $\phi$ tracer, we use the 545~GHz \planck\ CIB map~\cite{gnilc}.
For each data input, we either obtain existing corresponding simulations or generate them, as detailed in each subsection.
The simulated maps are processed identically as data in order to capture the effects of Wiener filtering and relative weighting between maps to properly propagate the uncertainties on the lensing template auto-spectrum.
A schematic of the lensing template pipeline is shown in Fig.~\ref{fig:pipeline_diagram}, which traces the three input branches through their respective processing steps to the final lensing $B$-mode template.

We impose the following multipole cuts as our fiducial setting: for the $E$ field we use $40 \leq \ell \leq 1500$, while for the $\hat{\phi}$ we use $24 \leq L \leq 1500$. For the CIB tracer, we apply $200 \leq L \leq 1500$. For the QE reconstruction inputs, we adopt the fiducial temperature and polarization multipole cutoffs from O26: $\ell_{\rm min}^{T/P} = 500$, $\ell_{\rm max}^T = 3500$, and $\ell_{\rm max}^P = 3000$. We explore the impact of varying these choices in Sec.~\ref{sec:analysis_choices}.
All maps are pixelized using HEALPix\footnote{\url{http://healpix.sourceforge.net} \cite{healpix, healpy}} at \texttt{Nside} = 2048.

\subsection{CMB Lensing Reconstruction as an Internal Tracer} \label{sec:qe_input}

We begin by describing lensing reconstructions obtained from CMB data using QEs, which exploit the mode coupling induced by gravitational lensing in the observed CMB temperature and polarization fields.

\subsubsection{CMB Lensing Reconstruction Formalism} \label{sec:qe_formalism}

Gravitational lensing introduces correlations between modes in the observed CMB that would otherwise be statistically independent in the absence of lensing. To leading order in the lensing potential, these correlations can be written as
\begin{equation} \label{eq:lensing_qe}
    \langle X_{\ell m} Y_{\ell' m'} \rangle_{\mathrm{CMB}} = \sum_{LM} (-1)^M
    \begin{pmatrix}
    \ell & \ell' & L \\
    m & m' & -M
    \end{pmatrix}
    f^{XY}_{\ell \ell' L} \phi_{LM},
\end{equation}
where $X,Y \in \{T,E,B\}$ denote the observed CMB temperature and polarization fields, $\phi$ is the lensing potential, and $f^{XY}_{\ell \ell' L}$ is the CMB lensing coupling coefficient. Here, $\langle \cdot \rangle_{\rm CMB}$ denotes an average over CMB realizations at fixed $\phi$.

QEs reconstruct the lensing potential $\phi$ from the lensing-induced mode couplings expressed in Eq.~\ref{eq:lensing_qe} using pairs of input CMB maps \cite{hu_okamoto_2002, okamoto_hu_2003}. 
In this work we consider both the global minimum variance (GMV) QE \cite{hirata_and_seljak, gmv}, which jointly incorporates all $T$, $E$, and $B$ information to construct a single combined estimator, and the sub-optimal quadratic estimator (SQE) \cite{hu_okamoto_2002, okamoto_hu_2003} for the polarization-only\footnote{Despite the nomenclature, as we explain later, for the combination of $EE$ and $EB$ estimators, the SQE is optimal.} estimator (hereafter PP), in which estimators are constructed individually for each map pair $XY$.
For the GMV estimator, we consider both the standard GMV and a profile-hardened GMV (hereafter GMVph), which is more resistant to extragalactic foreground biases.
All reconstructed lensing QEs used in this work are the products of the SPT-3G D1 lensing analysis, O26.

\paragraph{Standard GMV Estimator} \label{sec:gmv}

Below we briefly summarize the GMV formalism as introduced in \cite{hirata_and_seljak, gmv}. The GMV estimator for the lensing potential is given by
\begin{multline} \label{eq:gmv}
	\hat{\phi}^{\rm GMV}_{LM} = \frac{(R^{\rm GMV,\phi\phi}(L))^{-1}}{2} \times\\
    \sum_{\ell m} \sum_{\ell' m'} (-1)^M
    \begin{pmatrix}
	\ell & \ell' & L \\
	m & m' & -M \\
	\end{pmatrix}
    \overline{\bm{X}}^{T}_{\ell m} \bm{f}_{\ell \ell' L} \overline{\bm{X}}_{\ell' m'}.
\end{multline}
Here, $\bm{f}_{\ell\ell'L}$ is the matrix of lensing coupling coefficients \cite{okamoto_hu_2003}
\begin{equation}
    \bm{f}_{\ell\ell'L} =
    \begin{bmatrix}
    f^{TT} & f^{TE} & f^{TB} \\
    f^{TE} & f^{EE} & f^{EB} \\
    f^{TB} & f^{EB} & f^{BB}
    \end{bmatrix},
\end{equation}
and $\overline{\bm{X}}_{\ell m}$ is the inverse-variance-weighted observed CMB temperature and polarization fields
\begin{equation}
    \overline{\bm{X}}_{\ell m} =
    \begin{bmatrix}
    C_\ell^{TT} & C_\ell^{TE} & 0 \\
    C_\ell^{ET} & C_\ell^{EE} & 0 \\
    0 & 0 & C_\ell^{BB}
    \end{bmatrix}^{-1}
    \begin{bmatrix}
    T_{\ell m} \\
    E_{\ell m} \\
    B_{\ell m}
    \end{bmatrix},
\end{equation}
where the $C_\ell$ are total observed power spectra (signal plus noise) assuming $C_\ell^{TB} = C_\ell^{BT} = C_\ell^{EB} = C_\ell^{BE}=0$.
In idealized cases where the input maps are statistically isotropic, $R^{\rm GMV,\phi\phi}(L)$ enforces the estimator to be unbiased, $\langle \hat{\phi}^{\rm GMV}_{LM} \rangle_{\rm CMB} = \phi_{LM}$.
In O26, $R^{\rm GMV,\phi\phi}(L)$ is determined from the simulation-based response function, which accounts for effects such as masking and mode loss from anisotropic filtering.

Eq.~\ref{eq:gmv} shows how the GMV estimator jointly filters and combines the temperature and polarization fields to construct a minimum-variance quadratic estimator of the lensing potential. By incorporating information from all $T$, $E$, and $B$ modes simultaneously, the GMV reconstruction achieves a lower reconstruction noise than estimators constructed from individual field pairs \cite{gmv}.
Because this estimator does not explicitly mitigate foreground-induced mode couplings, the reconstructed lensing field can contain biases from non-Gaussian extragalactic foregrounds. Next we describe the GMVph estimator, which is designed to suppress some of these foreground-induced biases.

\paragraph{Profile-Hardened GMV Estimator}

QEs constructed from CMB temperature and polarization maps are sensitive not only to gravitational lensing but also to non-lensing sources of mode coupling. In particular, extragalactic signals such as the thermal Sunyaev-Zel'dovich (tSZ) effect, CIB, and radio sources can introduce additional mode couplings that can bias the reconstructed lensing potential \cite{osborne_hanson_dore_2014, mh_2018, crossilc_2023, van_engelen_2014}. These foreground-induced biases are especially significant for estimators that rely on temperature information, where foreground power dominates on small angular scales, which provide more mode pairs for each lensing $L$.

Profile-hardening is an estimator-level technique designed to mitigate such biases by modifying the quadratic estimator to subtract out a specified contaminant \cite{namikawa_hanson_takahashi_2013, osborne_hanson_dore_2014, sailer_schaan_ferraro_2020}. This is achieved by constructing an additional quadratic estimator $\hat{s}_{LM}$ that responds to the statistical anisotropies introduced by a field $s$ described by Poisson-distributed sources with specific profiles, and projecting it out from the lensing estimator.
A profile-hardened lensing estimator $\hat{\phi}^{\rm ph}$ can be obtained from
\begin{equation} \label{eq:prfhrd}
    \begin{bmatrix}
    \hat{\phi}^{\rm ph}_{LM} \\[2pt]
    \hat{s}^{\rm ph}_{LM}
    \end{bmatrix}
    = 
    \begin{bmatrix}
    R_L^{\phi \phi} & R_L^{\phi s} \\[2pt]
    R_L^{s \phi} & R_L^{s s}
    \end{bmatrix}^{-1}
    \begin{bmatrix}
    R_L^{\phi \phi} \hat{\phi}_{LM} \\[2pt]
    R_L^{s s} \hat{s}_{LM}
    \end{bmatrix},
\end{equation}
where the response functions $R_L^{a b} \equiv R^{a b}(L)$ quantify the response of estimator $a$ to field $b$.
The GMVph reconstruction used in this work follows the implementation described in O26, where the lensing estimator is hardened using a profile matching a modified tSZ power spectrum profile from \agora~\cite{agora}, and the source estimator is constructed for the $\bar{T}$ maps with corresponding $TT$ weights and responses.
For the choice of profile, noise levels of the input maps, and angular scales used, the normalized cross-response between $s$ and $\phi$ ranges between $\pm0.3$ with a zero crossing around $L$ of 2500. The responses are computed analytically without accounting for effects of anisotropic filtering, while the overall normalization is computed using simulations which accounts for mode loss (see O26 for details).
While profile-hardening increases the reconstruction noise relative to the standard GMV estimator, it reduces potential biases from foreground contamination, making it a more robust tracer of the lensing potential for use in constructing the lensing $B$-mode template. 
In this work, we take the GMVph lensing map combined with the CIB as our baseline $\phi$ tracer.

\paragraph{Polarization-Only SQE}

Although profile-hardening mitigates foreground contamination in the reconstructed lensing field, it relies on an assumed foreground profile and therefore cannot completely eliminate foreground-induced biases if the true contaminants differ from the assumed model. An alternative approach to avoid foregrounds is to construct a QE using only polarization maps. Because extragalactic foregrounds are largely unpolarized \cite{feng_holder_2019, khabibullin_2017}, polarization-only estimators at the SPT-3G D1 dataset noise levels are largely insensitive to these foregrounds.

We use a PP reconstruction constructed with the SQE formalism \cite{hu_okamoto_2002, okamoto_hu_2003}. In contrast to the GMV estimator described in Sec.~\ref{sec:gmv}, the SQE computes individual estimators from pairs of CMB maps, then linearly combines them with weights chosen to minimize the total variance.
Following the SQE formalism, the individual QEs can be written as
\begin{multline} \label{eq:sqe}
    \hat{\phi}^{XY}_{LM} = \frac{(R^{\rm XY,\phi\phi}(L))^{-1}}{2} \times\\
    \sum_{\ell m}\sum_{\ell' m'} (-1)^M
    \begin{pmatrix}
    \ell & \ell' & L \\
    m & m' & -M
    \end{pmatrix}
    f^{XY}_{\ell \ell' L}\,
    \overline{X}_{\ell m}\,\overline{Y}_{\ell' m'},
\end{multline}
where $X,Y \in \{E,B\}$ for the PP reconstruction, $\overline{X}$ and $\overline{Y}$ denote inverse-variance filtered polarization fields, and $R^{XY,\phi\phi}(L)$ enforces $\langle \hat{\phi}^{XY}_{LM} \rangle_{\rm CMB} = \phi_{LM}$.
The $EE$ and $EB$ estimators are then combined to form the final minimum-variance polarization-only estimate of the lensing potential:
\begin{equation} \label{eq:pp}
    \hat{\phi}^{\rm PP}_{LM} = \frac{\sum_{XY} \overline{\phi}^{ XY}_{LM}}{\sum_{XY} R^{XY,\phi\phi}(L)}
\end{equation}
where $XY \in \{EE, EB\}$ and $\overline{\phi}_{LM} = R^{\phi \phi}(L) \hat{\phi}_{LM}$ is the unnormalized lensing estimator. We note that a GMV-style joint combination of the $EE$ and $EB$ estimators reduces to this SQE combination, since the noise of the $EE$ and $EB$ estimators are uncorrelated ($C_{\ell}^{EB} = 0$); the two approaches are therefore equivalent for the polarization-only case.

Without temperature maps, we expect this estimator to have negligible contamination from extragalactic foregrounds, at the cost of higher reconstruction noise compared to the GMV estimators due to the absence of high signal-to-noise temperature-estimator lensing reconstruction.
This lensing template therefore provides a useful foreground-immune cross-check for the templates constructed using tracers that include temperature information.

\subsubsection{SPT-3G QE $\phi$ Data} \label{sec:qe_data}

\begin{figure}[t]
	\centering
	\includegraphics[width=\columnwidth]{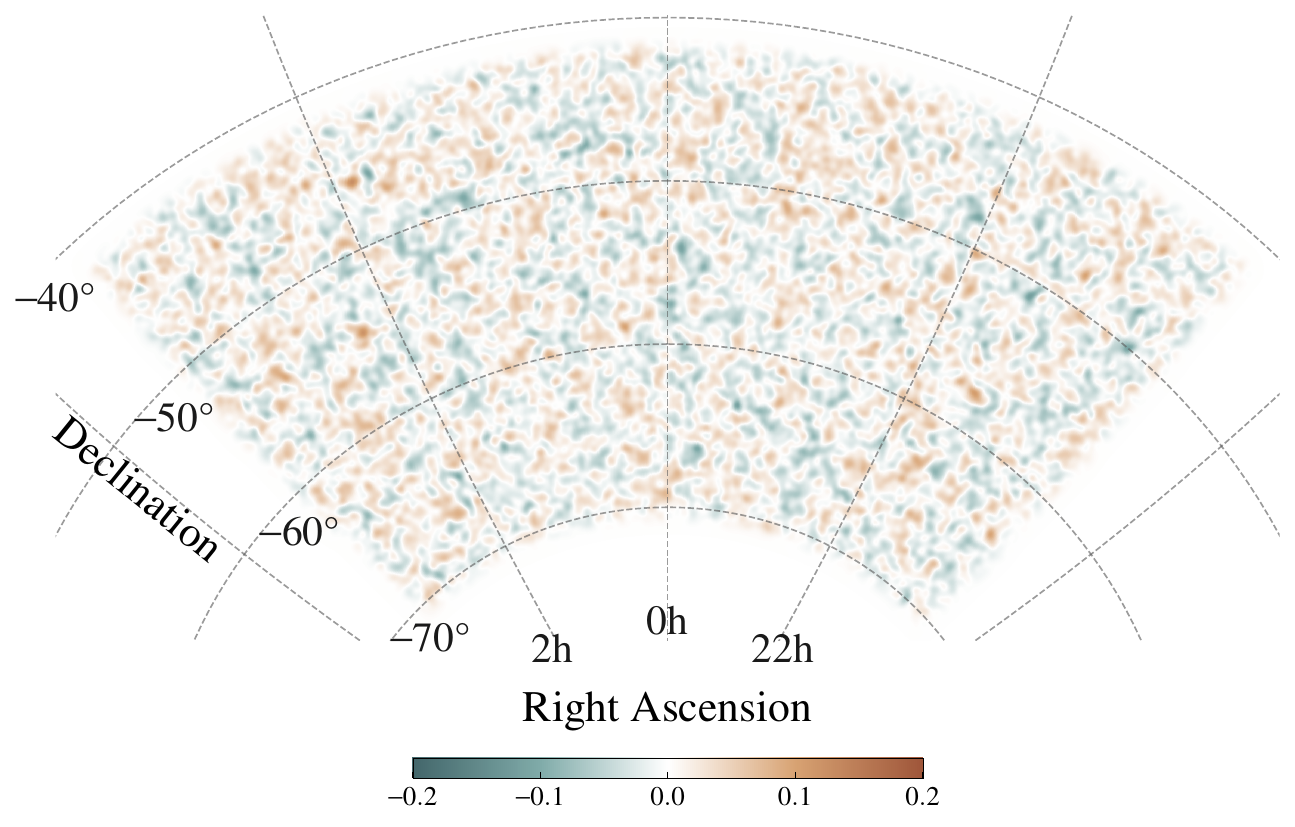}
	\centering \caption{The GMVph lensing convergence $\kappa$ map from data, smoothed with a 30 arcminute full width at half maximum~(FWHM) Gaussian beam for visualization. The GMVph $\kappa$ map is input to our baseline result.}
	\label{fig:data_kappa_map}
\end{figure}

The standard GMV, GMVph, and PP QEs are applied to the SPT-3G mid-$\ell$ map set in O26.
The mid-$\ell$ map set is optimized for high signal-to-noise measurements of sub-degree-scale temperature and polarization anisotropies. The maps are made from 95, 150, and 220~GHz observations, with full-depth $T$ white noise levels of 5.4, 4.4, and 16.2 $\mu{\rm K}$-arcmin in the three bands respectively, and 8.4, 6.6, and 25.8 $\mu{\rm K}$-arcmin for $Q$ and $U$ \cite{wei_maps}.
The mid-$\ell$ mapmaking applies a high-pass filter by removing low-frequency modes, including Legendre polynomial modes up to 30th order and sinusoids up to a scan-direction multipole cutoff of $\ell_x = 300$. This is done in order to suppress correlated $1/f$ atmospheric noise at the cost of losing large-scale modes. An anti-aliasing low-pass filter is also applied above $\ell_x = 13000$ to prevent high-frequency noise from aliasing into the map. For the baseline lensing reconstructions in O26, the $\ell_{\rm max}^T$ and $\ell_{\rm max}^{E/B}$ are 3500 and 3000, respectively, and the $\ell_{\rm min}$ is 500 for both temperature and polarization maps. Therefore, the modes in the lensing template below $\ell = 500$ are not biased by overlapping $B$ modes that enter the $\phi$ reconstruction~\cite{teng_2011, han_act_delens, carron_2017_internal_delens_plk, anton_internal_delens_biases}. We show the map of the GMVph-reconstructed $\kappa$ in Fig.~\ref{fig:data_kappa_map}.

How well the reconstructed $\phi$s are correlated with the true underlying lensing field, thus their sensitivity as a $\phi$ tracer, can be evaluated using their reconstruction noise power spectrum $N_L^{(0)}$.
Modes below $L \approx 400$ are signal-dominated (${\rm SNR} > 1$) for the standard GMV reconstruction, with the GMVph case having slightly degraded noise due to the hardening penalty.
The PP reconstruction has a higher $N_L^{(0)}$ (with $L \lesssim 250$ being signal dominated) because of the loss of temperature information, with negligible foreground-induced biases.

\begin{figure}[t]
	\centering
	\includegraphics[width=\columnwidth]{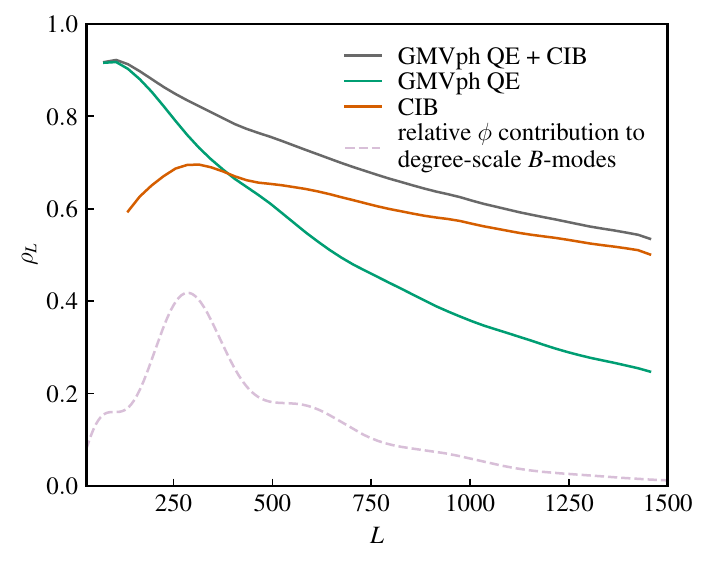}
	\centering \caption{The correlation coefficient $\rho_L^{\phi}$ between the tracer and true lensing convergence $\kappa$.
    Here we have also added $\sum_{\ell=[20,200]}(C_L^{\phi\phi}\frac{dC_\ell^{BB}}{dC_L^{\phi\phi}})/C_\ell^{BB}$ (scaled) for reference, which describes the relative contribution of $\phi$ power from each $L$ to the lensing $BB$ spectrum in the range $\ell = [20, 200]$.
    The QE-based internal tracer shows excellent correlation ($\approx$ 92\%; the highest to date) at low $L$ but worse correlations at higher $L$ due to noise, while the CIB tracer maintains a relatively constant correlation across all scales. The combined tracer improves the correlation over the full $L$ range, with the largest improvements at high $L$.
    }
    \label{fig:rhos}
\end{figure}

For QE-reconstructed $\phi$, the correlation between the lensing tracer and the true lensing potential (as defined in Eq.~\ref{eq:rho_phi}) becomes
\begin{equation} \label{eq:rho_phi_QE}
    \rho_L^{\phi,\rm QE} = \sqrt{\frac{C_L^{\phi^{\rm in}\phi^{\rm in}}}{C_L^{\phi^{\rm in}\phi^{\rm in}} + N_L^{\phi\phi,(0)}}}.
\end{equation}
This correlation is shown in Fig.~\ref{fig:rhos} for the GMVph QE, which reaches $\rho_L^{\rm QE} \approx 0.7$ near $L \approx 400$, consistent with the QE signal-dominated threshold.
The correlation reaches as high as $92\%$ at low $L$, the highest correlation to date, enabled by the highest-SNR-per-mode CMB lensing from O26.

Foreground robustness of the reconstructions is validated in O26 through estimator bandpower-difference tests. Comparing the bandpowers of GMVph against the PP reconstruction yields the probability-to-exceed (PTE) value of 0.90; the standard GMV versus PP comparison yields ${\rm PTE} = 0.80$, indicating consistency of the GMV and GMVph reconstructions against the PP reconstruction given the size of the difference-bandpower error bars.
In examining the difference-bandpower figure, O26 indicates that while the GMVph$-$PP points scatter around zero, the GMV$-$PP points are systematically below zero.\footnote{While almost all the difference-bandpower points below $L=1000$, where the test has statistical power, are below zero; the size of the difference error bars are large, so the PTE remains acceptable.} This is consistent with a small residual foreground bias at the $\lesssim 10\%$ level of the signal for $L \leq 1000$, as discussed in O26. These tests support the use of the GMVph estimator as a foreground-robust tracer, at a modest noise cost relative to the standard GMV.
We check the levels of foreground bias in the lensing template in Sec.~\ref{sec:foreground_bias}.

\subsubsection{SPT-3G QE $\phi$ Simulations} \label{sec:qe_sims}

We use the reconstructed lensing potential $\hat{\phi}$ for the GMV, GMVph, and PP estimators from the Gaussian and \agora\ simulation sets produced for O26. We briefly summarize these simulations and refer the reader to O26 for details.

The Gaussian simulation set consists of 500 independent realizations of the lensed primary CMB, along with Gaussian realizations of foregrounds, and sign-flip realizations of noise.
The simulated input sky maps are filtered and processed identically as the data maps.  
For each lensing estimator, this set of simulations is used to estimate the lensing mean-field and the simulation-based response, which are applied to the unnormalized data and simulation lensing maps to obtain unbiased lensing estimates.
In this work, these simulations are used to validate the lensing template construction pipeline and characterize the statistical uncertainties and delensing efficiencies in the absence of non-Gaussian foregrounds.

The \agora\ simulations~\cite{agora} are used to study the impact of non-Gaussian extragalactic foregrounds.
These simulations include lensed CMB along with lensed non-Gaussian realizations of extragalactic foregrounds: tSZ, kSZ, CIB, and radio sources. The foreground maps are processed with identical thresholds as data as follows: CIB and radio sources are masked using a single-pixel point source mask with a 6.0~mJy flux cut at 150~GHz, while the brightest tSZ clusters are masked and inpainted using a signal-to-noise threshold of SNR $> 10$ based on the map depth. Since there is only one full-sky realization of the \agora\ foregrounds, we cut out 10 patches with shapes that match the 1,500~deg$^2$ Main field.
The reconstructed $\hat{\phi}$ maps from \agora\ are constructed identically as the Gaussian simulations: from map processing to obtaining an unbiased $\phi$ estimate. 
We compute lensing templates using $\hat{\phi}$ from each estimator for all the Gaussian simulation realizations and the \agora\ patches.

\subsection{CIB as an External Tracer} \label{sec:cib_input}

Lensing reconstructions derived from CMB maps provide an unbiased, internal estimate of the lensing potential.
However, the reconstructed $\phi$ is not a perfect tracer of the underlying true $\phi$ due to non-zero reconstruction noise. For SPT-3G D1 CMB map noise levels, the GMV and GMVph reconstructions are signal-dominated for multipoles $L \lesssim 400$. 
However, lensing modes above $L$ of 400 contribute at a relevant level to degree-scale $B$ modes (see e.g., Fig.~\ref{fig:rhos}), motivating the use of external tracers that are more correlated with the true $\phi$ on complementary scales.

The CIB is a particularly well-suited external tracer. It is the integrated infrared emission of dusty star-forming galaxies across a broad range of redshifts, with a redshift distribution that peaks around $z \approx 2$ \cite{puget_1996, fixsen_1998}.
Because the CMB lensing potential is sourced by the integrated matter distribution along the line of sight with a redshift kernel that also peaks around $z \approx 2$, the CIB is moderately to strongly correlated with the true lensing field, at the $\gtrsim 60\%$ level across a broad range of scales relevant for generating degree-scale $B$ modes \cite{carron_lewis_2017, spt_bmode_2013, holder_2013, planck_lensing_cib_2013, act_lensing_cib_2015}. Moreover, the CIB maintains this correlation at angular scales where CMB lensing reconstructions become noise-dominated, making it complementary to lensing tracers derived from CMB maps alone \cite{agora}.

\subsubsection{545~GHz \planck\ CIB Map and Model Fitting} \label{sec:cib_data}

\begin{figure}[t]
	\centering
	\includegraphics[width=\columnwidth]{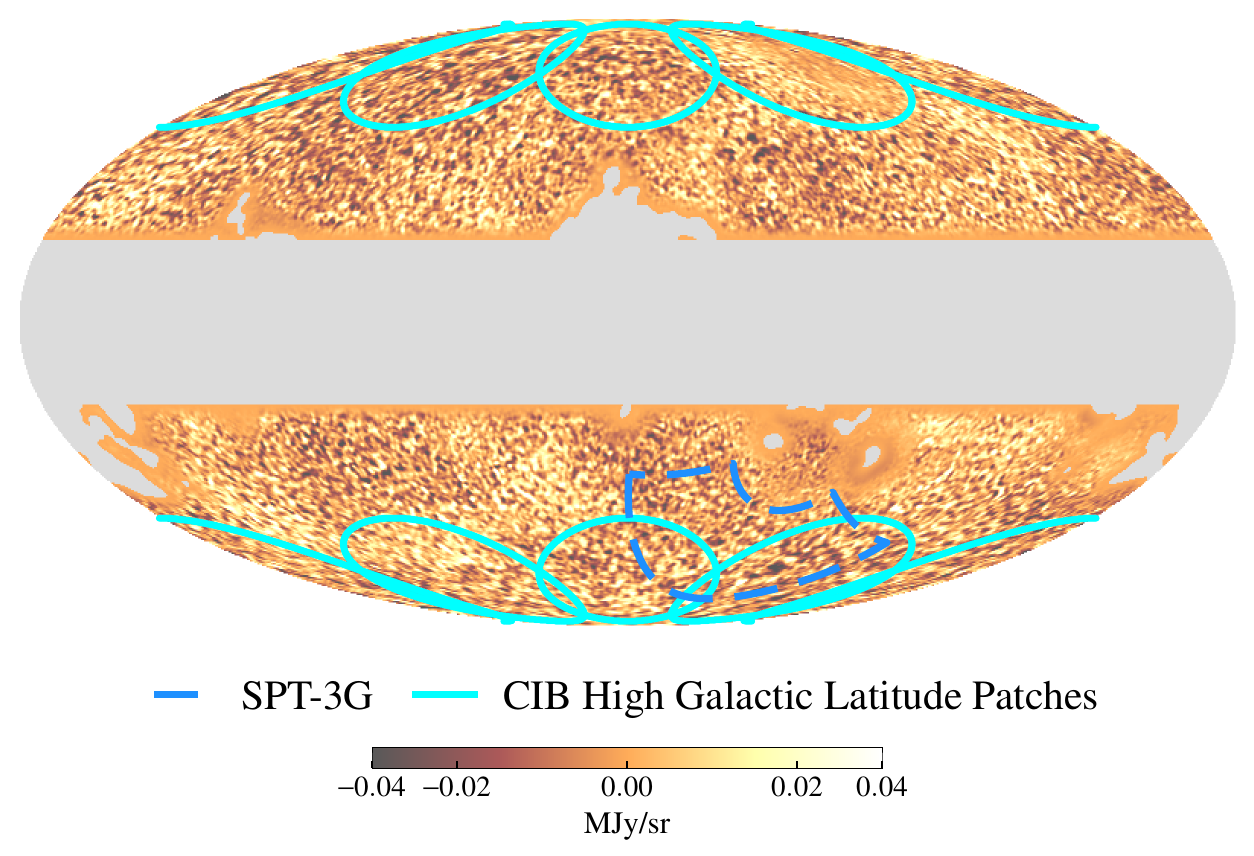}
	\centering \caption{The \planck\ 545~GHz GNILC CIB map, which we use as an external tracer of the lensing potential, shown in Mollweide projection in Galactic coordinates. The map is smoothed with a 50 arcminute FWHM Gaussian beam for visualization. The borders for the eight high-Galactic-latitude patches for computing the CIB auto-spectra, as well as the SPT-3G Main field are shown in green and blue, respectively.
    }
	\label{fig:data_cib_map}
\end{figure}

For the external $\phi$ tracer, we use the 545~GHz \planck\ CIB map constructed using the GNILC component separation method \cite{gnilc}.\footnote{There are CIB components solved at other frequencies (e.g., 353~GHz). They are highly correlated with the 545~GHz CIB map and do not add to the final $\kappa$-CIB correlation significantly. We also considered using CIB maps from~\cite{lenz_et_al} and~\cite{mccarthy_cib} and found reduced spatial overlap within the SPT-3G patch or lower correlation with the QE $\kappa$ than the 545~GHz \planck\ CIB map.}
GNILC is designed to separate Galactic thermal dust emission from CIB anisotropies by exploiting both their spectral and spatial differences. Because Galactic dust and the CIB have similar spectral energy distributions at \planck\ frequencies, frequency information alone is insufficient to disentangle the two components; GNILC supplements this with spatial information, using the fact that Galactic dust has a steeper power spectrum than the CIB.
The resulting CIB map covers approximately 57\% of the sky at high Galactic latitudes, shown in Fig.~\ref{fig:data_cib_map} in Galactic coordinates. In this work, we apply a minimum multipole cut $L_{\rm min}$ to the CIB map to avoid contamination from large-scale Galactic foreground residuals, adopting $L_{\rm min} = 200$ as our baseline.

To optimally combine and Wiener filter the CIB map with the QE $\phi$ tracer, we need to model how well the CIB correlates with the true lensing potential $\phi$ (or equivalently the lensing convergence $\kappa$). Specifically, we need $C_L^{I_{i}I_{i}}$ and $C_L^{I_{i}\kappa}$, where $i = {\rm CIB}$ for the inputs to Eq.~\ref{eq:combined_tracer_weights} for constructing the relative weights between CIB and QE $\phi$.

To obtain $C_L^{I_{\rm CIB}I_{\rm CIB}}$, we fit a power-law model $A(L/3000)^b + C$ to the CIB auto-spectra computed over the same 8 high-Galactic-latitude patches used in~\cite{bkspt_delens}.
We use the mean and scatter of the CIB auto-spectra from the 8 patches to obtain a more precise measurement than from the CIB within the SPT-3G patch alone. We avoid the lower Galactic latitudes to reduce Galactic foreground residuals.
The functional form of the fit without the constant $C$ was used in~\cite{manzotti_2017} and~\cite{spt_bmode_2013} to fit the CIB auto-spectrum.
The constant $C$ captures a shot-noise term that was employed in the CIB fit in~\cite{act_lensing_template}. The fitted model is a good description of both the mean CIB auto-spectrum across the 8 patches and within the SPT-3G Main field over the fiducial $L$ range between 200 and 1500.
Below $L$ of 200, the model predicts more CIB power than is measured from the spectra, which reduces the modeled correlation between the CIB and the true $\kappa$.
We do not modify the model to improve the fit below $L$ of 200 because (1) we do not use them in our fiducial setting, and (2) we would like to suppress contributions from the CIB at $L < 200$ even if we include them (e.g., in data-difference tests) to avoid biases from residual Galactic foregrounds in the CIB map.

For $C_L^{I_{\rm CIB}\kappa}$, we model the CIB using the single-SED model of~\cite{hall_2009} with a CIB emissivity peak at redshift $z_c = 2$ and a broad redshift kernel of $\sigma_z = 2$ and CMB lensing assuming $\Lambda$CDM. We use the Limber approximation to compute the model cross-spectrum between CIB and lensing, and fit the linear bias term in the CIB model to the cross-spectrum between the \planck\ 545~GHz CIB map and the SPT-3G PP lensing map.
We use the SPT-3G PP map for this fit because it has negligible foreground biases, which otherwise can bias the cross-spectrum measurement via bispectra sourced by combinations of CIB and tSZ.
As a comparison, this fit agrees to within $\approx 1\%$ with a fit of the same model to the mean cross-spectrum between the 545~GHz CIB map and the \planck\ PR4 lensing maps across the 8 patches, a difference that is consistent within 1$\sigma$ of the fit uncertainty.

We use these model fits for both the weights to combine the CIB with the QE $\phi$ and for constructing the CIB simulations. 
For the weights, the fits are used for the  $C_L^{I_{\rm CIB}I_{\rm CIB}}$, $C_L^{I_{\rm CIB}\kappa}$, and $C_L^{I_{\rm CIB} I_{\hat{\kappa}}}$ terms in Eqs.~\ref{eq:rho_ij} and~\ref{eq:rho_ik}, where $\hat{\kappa}$ denotes the measured lensing convergence.
For the CIB simulations, they are used for constructing CIB noise realizations, discussed in the next section.
We show that these model fits are reasonable descriptions of the data in the context of the lensing template by comparing the data with the simulation mean lensing template auto-spectrum in Sec.~\ref{sec:lensing_template_spectra}.

We compute the correlation coefficient $\rho_L^{\phi}$ for this CIB map as  $C_L^{I_{\rm CIB}I_{\hat{\kappa}}} / \sqrt{C_L^{I_{\rm CIB}I_{\rm CIB}} C_L^{\kappa\kappa} }$ using the fits obtained.
We show the resulting correlation coefficient $\rho_L^{\phi}$ for the CIB $\phi$ tracer, GMVph, and their combination in Fig.~\ref{fig:rhos}.
As expected, the correlation coefficient for the QE-only tracer decreases at high $L$, where the CMB lensing reconstruction becomes increasingly noise-dominated. In contrast, the CIB-only tracer exhibits a stabler correlation across all angular scales, and exceeds the QE-only correlation at $L \gtrsim 400$.
Combining the two yields a tracer with consistently high correlation, $\rho_L^{\phi} \approx 60$--$90\%$, across the full $L$ range of $\phi$ input to the lensing template. 

\subsubsection{CIB Simulations} \label{sec:cib_sims}

To generate simulated CIB realizations that have the appropriate correlation with the input lensing field, we follow the procedure used in \cite{bkspt_delens}, which requires an auto-spectrum of the CIB map and cross-spectum between the CIB map and the lensing field.
We obtain these auto- and cross-spectra by fitting to the measured CIB auto-spectrum and CIB-$\kappa$ cross-spectrum, as discussed in the previous section.
For the simulation generation method below, we denote the CIB map by $I_{\rm CIB}$, with $C_L^{I_{\rm CIB}I_{\rm CIB}}$ representing its auto-spectrum and $C_L^{I_{\rm CIB}\kappa}$ its cross-spectrum with $\kappa$.

The noise component of the simulated CIB, $I^\textnormal{noi}_{LM}$, is drawn from the spectrum
\begin{equation} \label{eq:I_noi}
    N_L = C_L^{I_{\rm CIB}I_{\rm CIB}} - \frac{(C_L^{I_{\rm CIB}\kappa})^2}{C_L^{\kappa\kappa}},
\end{equation}
and is generated as Gaussian realizations in harmonic space using \texttt{synalm} \cite{healpix, healpy}. 
Here, $C_L^{\kappa\kappa}$ is the fiducial lensing power spectrum from \camb.\footnote{\url{https://camb.info} \cite{camb}}
Then, the signal part is constructed as
\begin{equation} \label{eq:I_sig}
    I^\textnormal{sig}_{LM} = \frac{C_L^{I_{\rm CIB}\kappa}}{C_L^{\kappa\kappa}} \kappa^{\textnormal{in}}_{LM}.
\end{equation}
We transform the harmonic signal and noise coefficients to pixel space, and
the simulated CIB tracer is given by
\begin{equation}\label{eq:cib_sim_combo}
    I_{\rm CIB} = I^\textnormal{sig}_{\rm CIB} + I^\textnormal{noi}_{\rm CIB}.
\end{equation}
We then apply the apodized boundary mask\footnote{Since the boundary of the lensing mask encloses that of the $\mathbb{C}^{-1}$ mask (for Wiener filtering the $E$ modes; see Sec.~\ref{sec:E_mode_filtering}), the $Q/U$ polarizations are not lensed by edge $\phi$ modes that are partially reconstructed.} for the lensing maps in O26 to form the input to combine with the QE-reconstructed $\phi$ tracers.
The simulated CIB realizations generated using this procedure preserve the measured correlation between the CIB and the lensing field while allowing independent realizations of the noise component. We use these simulations to validate the analysis pipeline and estimate statistical uncertainties associated with the use of the CIB tracer.

For the \agora\ simulations, we use \agora\ CIB component simulations synthesized at 545~GHz, co-located with the 10 patches used for QE reconstruction.
To ensure that the CIB--lensing cross-correlation coefficient $\rho_{I_{\rm CIB}\kappa}$ (Eq.~\ref{eq:rho_ik}) is consistent between the \agora\ and Gaussian CIB simulations, we add Gaussian noise to the \agora\ CIB maps at the level of
\begin{equation}
    N_L^{\agora} = C_L^{I_{\rm CIB}I_{\rm CIB}} \left(\frac{C_L^{I_{\rm CIB}\kappa,\ \agora}}{C_L^{I_{\rm CIB}\kappa}}\right)^2 - C_L^{I_{\rm CIB}I_{\rm CIB},\agora}.
\end{equation}
This elevates the \agora\ CIB auto-spectrum while leaving the CIB--lensing cross-spectrum unchanged, such that $\rho_{I_{\rm CIB}\kappa}$ matches between the two simulation sets.
Matching correlations ensures that the lensing $B$ power is consistent between the \agora\ set and the Gaussian simulation set in the combined $\phi$ tracer case such that differences can be attributed to foregrounds.

\subsection{$E$ Map Input} \label{sec:E_input}

The $E$ modes for the lensing template are derived from the SPT-3G low-$\ell$ map set \cite{lowell_bb}. We first describe this map set and its simulations, followed by the Wiener-filtering formalism used to process the $E$ modes as input to the lensing template.

\subsubsection{SPT-3G Low-$\ell$ Map Set for $E$ Modes} \label{sec:E_mode_data}

The SPT-3G D1 low-$\ell$ map set is designed to preserve large-scale polarization modes relevant for degree-scale $B$-mode analyses~\cite{lowell_bb}. 
In particular, unlike in the mid-$\ell$ map set,
not all modes below angular multipoles of $\ell \approx 300$ are filtered away.
Since more than 50\% of the lensing $B$ modes on degree scales ($20 \leq \ell \leq 200$) are sourced by $E$ modes below $\ell \approx 420$, retaining these large-scale $E$ modes is important for constructing a high-fidelity lensing template.

We briefly summarize the steps taken in making the low-$\ell$ map set and refer the reader to~\cite[][hereafter Z25]{lowell_bb} for details.
In contrast to the mid-$\ell$ maps, which rely on aggressive high-pass filtering to suppress noise from atmospheric fluctuations, the low-$\ell$ processing includes dedicated mitigation of polarized atmospheric noise and optimization of low-$\ell$ noise while retaining large-scale signal. 
A milder high-pass filter is applied by subtracting a 10th-order polynomial from each scan, suppressing $1/f$ noise without removing the degree-scale modes of interest.
A low-pass filter is also applied at a lower cutoff of $\ell_x = 3000$ to reduce aliased noise due to the finite pixel resolution at \nside = 2048.
Additionally, the timestreams from pixel pairs are gain-matched using elevation slews, and the pair-differenced timestreams are used for calculating the weights of each pair.
Besides the timestream-level operations, the low-$\ell$ map set optimizes $\ell < 200$ noise at the map level by (1) subtracting a coadd-subtracted scaled copy of the 220~GHz $Q$ map from the 150~GHz $Q$ map to reduce polarized atmospheric noise and by (2) weighting the observations using noise in the $\ell=50$--$250$ range when forming the full co-added map.

In this work, we use the 95~GHz and 150~GHz polarization maps from this low-$\ell$ map set to construct the Wiener-filtered $E$ modes used as input to the lensing $B$-mode template.
In the following subsections, we describe the map calibration and the 95~GHz and 150~GHz map combination procedure. 

\paragraph{Map Calibration}

While the input observation maps are identical, some of the map calibration steps differ from those applied to the map set that produced the $B$-mode spectrum measurement in Z25.
Specifically, in Z25, temperature-to-polarization (T-to-P) leakage and polarization calibration were corrected at the spectrum level after the observations from the four sub-fields were coadded to form the full map.
In this analysis, however, these corrections must be applied at the map level, as the lensing $B$-mode template requires calibrated, leakage-corrected maps for subsequent delensing analyses.

For the map calibration steps, we largely follow the procedures for the mid-$\ell$ map set detailed in~\cite{wei_maps}.
We compute the T-to-P leakage coefficients and subtract the monopole leakage from the $Q/U$ map in an identical manner to the mid-$\ell$ map set.
We find the leakage coefficients to be smaller than those in the mid-$\ell$ map set: the 95~GHz coefficients are between 0.1 and 0.2\% for $Q$ and $U$ across all four sub-fields; the 150~GHz coefficients are negligible ($<$ 0.1\%).
The small T-to-P leakage is expected because this leakage is dominated by gain mismatch between detector pairs and is mitigated by the gain matching implemented for this map set.

After correcting for T-to-P leakage, we apply per-sub-field polarization calibration factors ($P_{\rm cal}$) following the temperature calibration calculations for the mid-$\ell$ map set.
We compute $P_{{\rm cal}, i}^{150}$ by taking the ratio of the cross-spectrum of the SPT 150~GHz map with \planck{} 143~GHz map and the cross-spectrum between two halves of SPT 150~GHz maps for sub-field~$i$. The inverse-covariance-weighted-average ratio between $\ell$ of 400 and 1000 determines the $P_{\rm cal}$ value of each sub-field.
The calibration factors for the 95~GHz map are determined internally using SPT-3G 95~GHz and 150~GHz maps and chained to $P_{\rm cal}^{150}$.
Specifically, $P_{{\rm cal}, i}^{95} = P_{{\rm cal}, i}^{150} / \eta_{{\rm cal}, i}^{95}$ with  $\eta_{\rm cal, i}^{95}$ being the ratio of the cross-spectrum of the SPT 150~GHz map and the SPT 95~GHz map and the cross-spectrum between two halves of SPT 150~GHz maps for sub-field~$i$, similarly average-weighted.

We coadd these per-subfield calibrated $Q/U$ maps to form the full-field map and then apply an overall rotation angle correction.
We check that the $\Delta\psi_{\rm cal}$ values derived using the mid-$\ell$ maps null the $EB$ cross-spectrum of this low-$\ell$ map set and apply identical $\Delta\psi_{\rm cal}$ corrections.
The per-subfield calibration factors have statistical uncertainties of $1$--$2\%$, with the smaller subfields having larger uncertainties, dominated by \planck's polarization map noise. So we additionally derive full-map calibration factors by comparing to bandpowers from the mid-$\ell$ map set.
We compare the debiased mid-$\ell$ 95~GHz and 150~GHz $EE$ bandpowers\footnote{We apply the best-fit $T_{\rm cal}$ and $E_{\rm cal}$ parameters derived from fitting the SPT-3G D1 $TT/TE/EE$ bandpowers across all three bands to the $\Lambda$CDM model to the debiased bandpowers.} with the beam- and transfer-function-corrected low-$\ell$ $EE$ bandpowers and apply a 1.9\% and a 2.3\% correction to the 95~GHz and the 150~GHz maps, respectively.

\paragraph{95 and 150~GHz Map Combination} \label{sec:E_mode_data_map_combination}

Prior to the Wiener filtering step, we construct minimum-variance combinations of the 95 and 150~GHz $Q/U$ maps.
Assuming negligible foregrounds,\footnote{With an inverse-noise-variance weight, the dust power on this patch of the sky given by the best-fit dust model from the BICEP measurement~\cite{bk18} is subdominant to the noise power in the $\ell$ range considered ($\ell$=[40, 1500]).} the minimum-variance weights are given by the inverse noise variance of the maps:
\begin{equation}\label{eq:invnoi}
    m_{\rm comb} = \frac{ \sum w^{\nu} m^{\nu}}{\sum w^{\nu}}, 
\end{equation}
where $m^{\nu}$ denotes the map at frequency $\nu$, $\nu \in [95, 150]$~GHz, and $w^{\nu} = 1/N_{\ell}^{\nu}$.
$N_{\ell}$ is computed by averaging 500 beam- and transfer-function-corrected noise power spectra.
The maps and noise realizations are masked using a variant of the mask used in Z25\footnote{We use the mask used during development of Z25~\cite{lowell_bb}, which contains apodized point source masks for sources with temperature flux above 10~mJy in the 150~GHz band, as well as an apodized boundary.} prior to computing the power spectrum.
The combined map is signal-dominated in $E$ modes for multipoles below $\approx$~2300.

\subsubsection{SPT-3G Low-$\ell$ Map Set $Q/U$ Simulations} \label{sec:E_mode_sims}

We create 500 signal simulations by mock-observing simulated lensed CMB skies identically as real data maps at 95 and 150~GHz of the low-$\ell$ map set.
The input skies are identical to the set used in O26 in order for the lensing $B$ template to trace the input lensing $B$ modes.
To form a complete mock sky, we add to the mock-observed $Q/U$ signal maps noise realizations from their respective frequency bands.
The sign-flip noise realizations are generated as described in~\cite{wei_maps, lowell_bb}, where the coadd of all observations is subtracted from each individual observation to remove the signal, individual observations are multiplied by a random sequence of $+1$ and $-1$, and the signed observations are then averaged according to their observation weights.\footnote{In principle, the noise realizations for the low-$\ell$ and mid-$\ell$ map sets could be constructed so that the sign-flip sequence for common observations are identical for each sky seed. We test this for a subset of the simulations and observe negligible difference in the lensing template auto-spectrum, its cross-spectrum with the input $B$, and their respective variances. We thus proceed with using independent sign-flip sequences for the low-$\ell$ map set in order to achieve more even weighting across the two halves of the observations forming noise realizations.}
As done on data, this set of mock $Q/U$ skies at 95 and 150~GHz are inverse-noise-variance combined prior to $\mathbb{C}^{-1}$ filtering, as part of the $E$ Wiener filter, introduced in the next subsection.

Besides serving as input to the simulated $E^{\rm WF}$ for generating simulated lensing templates, this set of simulations is used to model the data in the $\mathbb{C}^{-1}$ step and as weights when combining the  95 and 150~GHz polarization maps.
For the $\mathbb{C}^{-1}$ step, we construct a model for the transfer function and the noise spectrum, both in $\ell, m$ space (not $m$ averaged).
We estimate the 2D transfer function by taking the square root of the average of the ratio of the mock-observed and the input non-$m$-averaged-$E$-mode spectrum across 500 realizations.
We compute the model 2D noise spectrum by averaging non-$m$-averaged noise spectra for each frequency band.
The 2D transfer functions and noise spectra for each band are inverse-noise-variance combined, and the combined noise spectrum is smoothed using a Gaussian filter prior to passing them to the $\mathbb{C}^{-1}$ filter.
As for the inverse-noise weights used for combining the data maps (Eq.~\ref{eq:invnoi}), the transfer functions and the noise spectra are 1D objects constructed for 95 and 150~GHz.
These noise spectra are computed by averaging 500 noise spectra for each band, debiased using the respective beam and 1D transfer function estimated similarly as their 2D counterparts but with $m$ averaging.

\subsubsection{Wiener-Filtered $E$ Map} \label{sec:E_mode_filtering}

We form the $E^{\rm WF}$ input to the $B$ template by multiplying the inverse-variance-filtered $E_{\ell m}$, $\bar{E}$, by the signal variance $C^{EE}_{\ell}$. 
The inverse-variance filtering on the input polarization $Q/U$ maps are performed using the $\mathbb{C}^{-1}$ approach, commonly employed in CMB lensing analyses~\cite[e.g.,][]{spt_lensing_20192020, pan_lensing, sptpol_lensing}. 

In brief, the total variance $\mathbb{C}$ of the input maps is modeled to have a harmonic-space component $\mathbb{S}_{\ell m}$ and a pixel-space component $\mathbb{N}_{\rm flat}(\hat{n})$. 
In operator form, 
\begin{equation}
\mathbb{C} = T \mathbb{S} T^{\dagger} + \mathbb{N}_{\rm flat},
\end{equation}
with $T$ denoting a transfer function.
The harmonic component includes both the CMB lensed power spectra and the non-flat, residual noise:
\begin{equation}
\mathbb{S}_{\ell m}  = C_{\ell}^{\rm XX} + N_{\ell m}^{\rm XX, res}
\end{equation}
where $C_{\ell}$ denotes the theory spectra from \camb\ evaluated at the fiducial cosmology, ${\rm X \in [E, B]}$, and the $\ell m$ subscripts denote that the objects depend on both $\ell$ and $m$, which is important since the filtering operations lead to non-isotropic maps. 
The residual noise term $N_{\ell m}^{\rm XX, res}$ is determined by subtracting the variance of a white-noise level of $5/\sqrt{2}$~\ukam\ from the 2D pseudo noise power spectra and then debiased with the 2D transfer function, described previously.
The noise level is chosen to match the noise power in the filtered maps at $\ell \approx 3000$.
The pixel-space component $\mathbb{N}_{\rm flat}(\nhat)$ is constructed by filling HEALPix pixels \cite{healpix, healpy} with noise variance values corresponding to a $5/\sqrt{2}$~\ukam\ noise level. 
The $\mathbb{C}^{-1}$ mask, constructed by taking the union of the mask used for combining the 95 and 150~GHz maps and the $\mathbb{C}^{-1}$ mask used on polarization maps in O26,\footnote{The $\mathbb{C}^{-1}$ mask in O26 additionally identified source locations using their polarization fluxes.}
is multiplied to $\mathbb{N}_{\rm flat}^{-1}(\nhat)$ so that masked pixels have infinite variance and are downweighted. 

We compute the inverse-variance-filtered fields $\bar{X}$ by solving~\cite{smith_2007}
\begin{equation}
\bar{X} = \mathbb{S}^{-1} \!\left[ \mathbb{S}^{-1}+T^{\dagger} \mathbb{N}_{\rm flat}^{-1} T \right]^{-1}
T^{\dagger}\mathbb{N}_{\rm flat}^{-1}d
\end{equation}
using preconditioned conjugate-gradient descent, where $X$ denotes $E/B$ and $d$ is the input $Q/U$ maps.
The Wiener-filtered $E^{\rm WF}_{\ell m}$ is given by $C^{EE}_{\ell} \bar{E}_{\ell m}$, and has a harmonic-space mask, which removes a narrowband contamination noted in~\citet{wei_maps}, applied prior to passing to the lensing template estimator.

\section{Results} \label{sec:results}

\begin{figure*}[t]
	\centering
	\includegraphics[width=\textwidth]{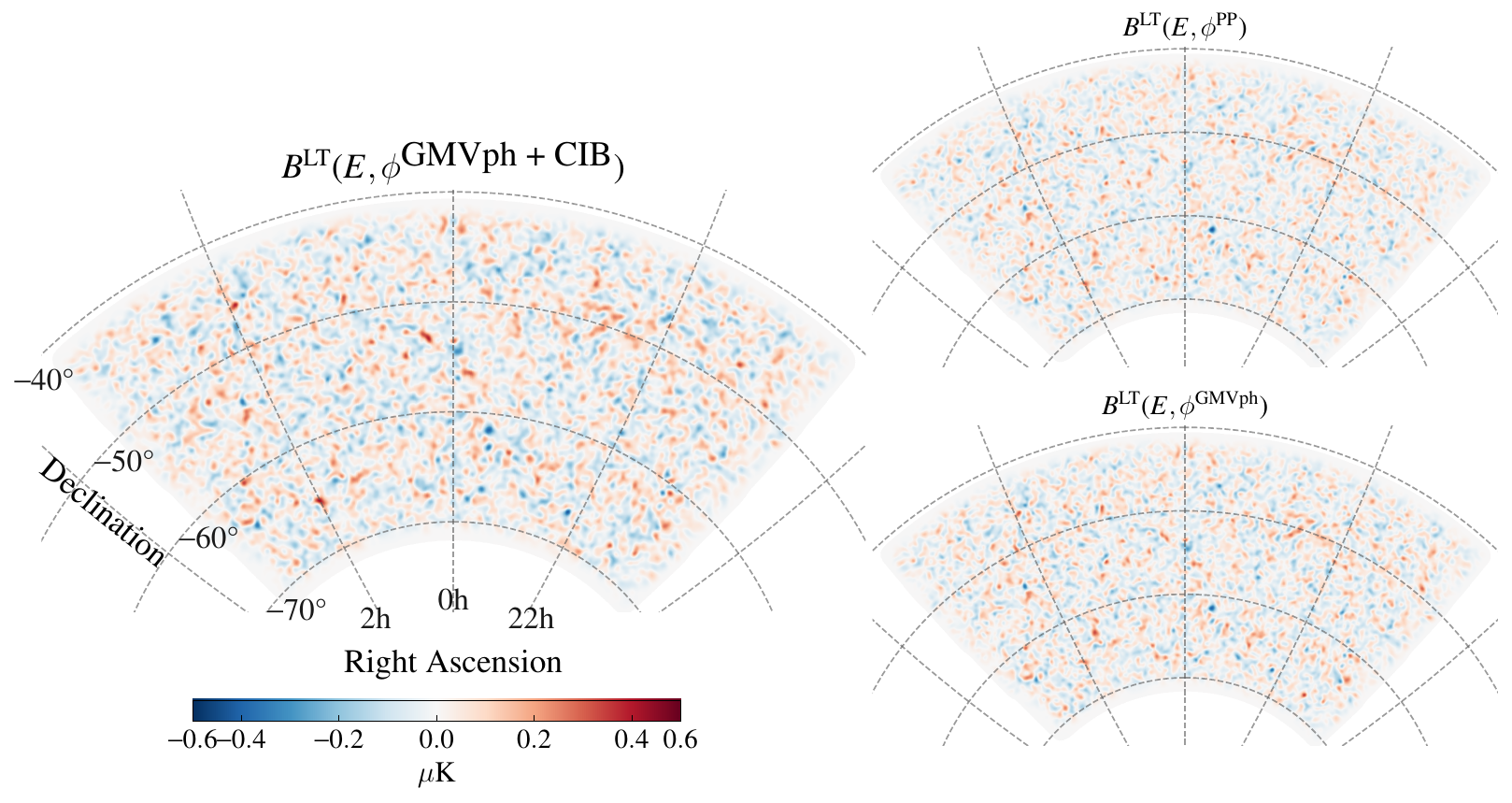}
	\centering \caption{The lensing template map from data, constructed using PP (top right), GMVph (bottom right), and GMVph + CIB (left) as the $\phi$ tracers, smoothed with a 30 arcminute FWHM Gaussian beam for visualization. All maps share the same colorbar. The relative ordering of amplitudes is determined by each tracer's correlation with the true lensing field: the GMVph + CIB combination has the highest $\rho_L^{\phi}$ and recovers the most lensing $B$-mode power, while PP, with the lowest $\rho_L^{\phi}$, recovers the least.}
	\label{fig:data_LT_map}
\end{figure*}

In this section we present our main results.
We begin by examining the lensing $B$-mode template maps constructed from data and comparing their amplitudes across different choices of lensing tracers. We then compare the template power spectra measured from data to expectations from Gaussian simulations, providing a validation of the signal and noise modeling used throughout the analysis.
Next, we quantify the delensing efficiency achieved by each $\phi$ tracer and study the robustness of these results to variations in the multipole cut analysis choices. We then assess the impact of extragalactic foregrounds, and finally discuss the implications for $\sigma(r)$.

Throughout this section, we restrict our lensing template spectra to $\ell \leq 500$, as using $B$ modes at smaller scales in the template would introduce a bias from the $B$ modes shared between the $EB$ lensing reconstruction and the template \cite{teng_2011, han_act_delens, act_lensing_template, polarbear_delensing_2019,  carron_2017_internal_delens_plk, anton_internal_delens_biases}.

\subsection{Lensing Template Maps} \label{sec:lensing_template_maps}

Fig.~\ref{fig:data_LT_map} shows the lensing template maps constructed from data using the PP, GMVph, and GMVph + CIB tracers. All three maps trace the same underlying lensing $B$-mode structure, reflecting the fact that they share the same $E$-mode input and are reconstructed using lensing fields that trace the same large-scale structure. The amplitude of the template varies across the tracers, with the GMVph + CIB tracer, the baseline in this work, recovering the most lensing $B$-mode power, and the PP tracer recovering the least.

To understand this relative ordering of amplitudes, recall from Eq.~\ref{eq:B_template} that the lensing template is constructed by convolving the Wiener-filtered $E$ and Wiener-filtered $\phi$ tracer. Since the same $E$ mode is used in all cases, differences in template power are driven primarily by the $\phi$ tracer.
The $\phi$ Wiener filter $W^{\phi}_L$, as defined in Eq.~\ref{eq:qe_wiener_filter} for the QE-only case, can be rewritten in terms of the correlation coefficient $\rho_L^{\phi}$ (defined in Eq.~\ref{eq:rho_phi}) as
\begin{equation}
    W^{\phi}_L = \left(\rho_L^{\phi}\right)^2
\end{equation}
assuming that the tracer noise is uncorrelated with the true $\phi$.
For the multi-tracer case, the analogous role is played by the combined-tracer weights $c_i$ (Eq.~\ref{eq:combined_tracer_weights}), which perform Wiener filtering and tracer combination in a single step.
In both cases, the amplitude of the Wiener-filtered $\phi$ map increases with the correlation between the tracers and the true lensing field, which in turn increases the power in the resulting lensing  $B$ template.
Thus, tracers combining QE and CIB which have a higher $\rho_L^{\phi}$ (Fig.~\ref{fig:rhos}) retain more signal after Wiener filtering, and therefore generate lensing templates with more $B$-mode power.

\subsection{Lensing Template Spectra} \label{sec:lensing_template_spectra}

\begin{figure*}[t]
	\centering
	\includegraphics[width=\textwidth]{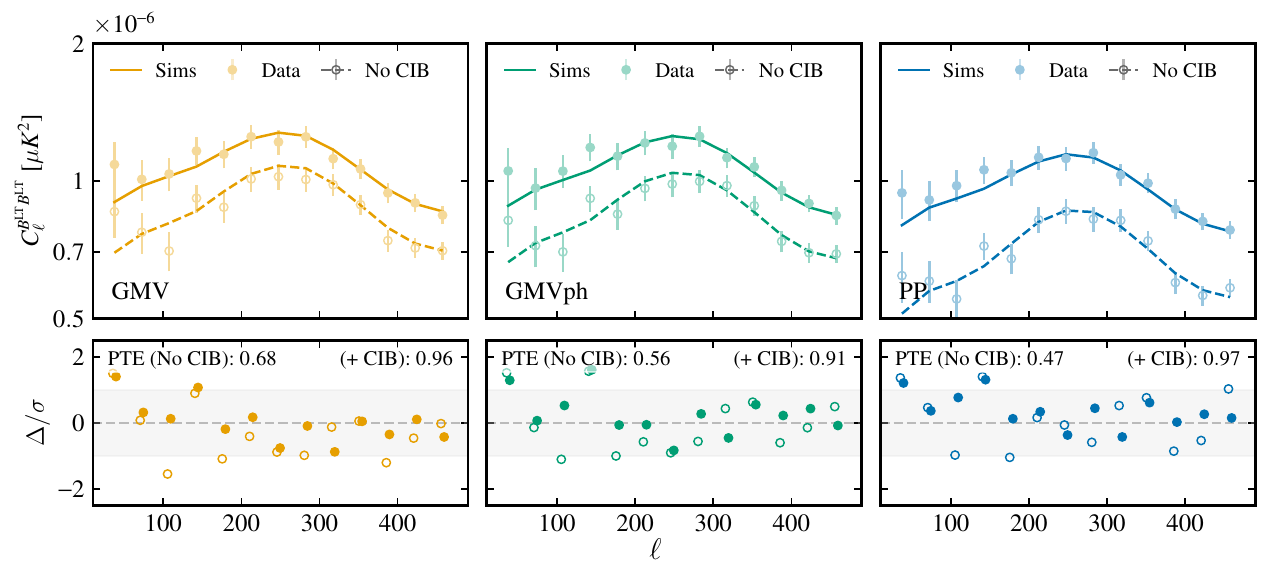}
	\centering \caption{The lensing template auto-spectra from data (lighter colored points), compared to the Gaussian-simulation mean (darker colored lines), constructed using the standard GMV (left panel), GMVph (middle panel), and PP (right panel) as the $\phi$ tracer. We include cases with (solid lines, filled points) and without (dashed lines, open points) the CIB tracer. The error bars are the standard deviation across all Gaussian simulations.
    The bottom subplots show the difference between the data and simulations, normalized by the standard deviation.
    The data auto-spectra are consistent with the Gaussian simulation means across all tracer choices.
    For reference, the fiducial lensed $BB$ spectrum from \camb\ is approximately $2\times10^{-6}$ $\mu{\rm K}^2$ across the plotted range; we do not overplot it to avoid expanding the $y$-axis and compressing the template spectra.
    }
	\label{fig:data_LT_specs}
\end{figure*}

We next compute the lensing template auto-spectra measured from data, compared to the Gaussian-simulation mean. We use 499 Gaussian simulations, as described in Sec.~\ref{sec:inputs}, which contain no non-Gaussian foregrounds and therefore provide a straightforward test of the pipeline.
The power spectra are computed using \texttt{PolSpice}, on masked maps using the $\mathbb{C}^{-1}$ filtering mask.
Fig.~\ref{fig:data_LT_specs} shows the resulting power spectra for each tracer choice.

In all tracer cases, only a fraction of the input lensing $B$-mode power is recovered. Perfect reconstruction is not expected due to tracer noise, masking and apodization, and filtering, all of which suppress or mix modes. The relative ordering of the estimators is consistent with expectations: the standard GMV reconstruction yields the highest template power, followed by the GMVph case, while the PP reconstruction has the lowest power due to its higher reconstruction noise. In all QE cases, the addition of the CIB improves the correlation with the true lensing signal, leading to lower residual lensing power, consistent with what was discussed in Sec.~\ref{sec:cib_data}.
The relative delensing efficiencies are also consistent with the ordering of amplitudes in the maps shown in Fig.~\ref{fig:data_LT_map}.

We also see from Fig.~\ref{fig:data_LT_specs} that the data auto-spectra are consistent with the Gaussian-simulation means across all tracer choices, with PTEs ranging from 0.47 to 0.97.
For $\ell < 200$ in the with-CIB cases, there appears to be a slight positive excess. These are unsurprising statistical fluctuations given the correlations between neighboring bins.
To verify this, we compute a PTE derived from a $\chi$ test by summing the per-bin residuals normalized by their standard deviations and comparing to the same statistic computed on simulations. The resulting $\chi$ PTEs range between 0.36 and 0.95 across all tracer choices for the entire $\ell$ range, and between 0.13 and 0.96 for the first five bins.
The agreement between the data and Gaussian simulations indicates that the Gaussian simulations provide an accurate description of the signal and noise properties in the data.
Since the same signal and noise models are used to construct the Wiener filter weights applied to the data, the fact that they accurately describe the data implies that the weighting and filtering on the data are also near-optimal. We further discuss the near-optimality of the Wiener filtering in Appendix~\ref{sec:filtering_and_weighting}. It is also important that the Gaussian simulations accurately reflect the data because they are used to model lensing template cross-spectra with other bands in the $r$ inference during delensing.

For the standard GMV case (open circles), the data-minus-simulation-mean points lie systematically lower compared to both the GMVph and PP cases, suggesting a low level of foreground-induced bias in the GMV temperature reconstruction. We quantify this in Sec.~\ref{sec:foreground_bias} with  difference tests between estimators (Fig.~\ref{fig:agora_data_spec_diff}).

\subsection{Delensing Efficiency}

\begin{figure}[t]
	\centering
	\includegraphics[width=\columnwidth]{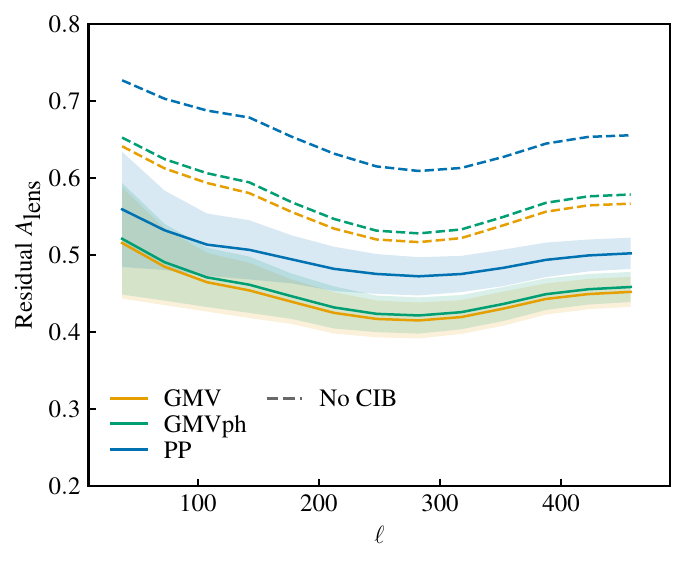}
	\centering \caption{Residual lensing amplitude $A_{\rm lens}^{\rm res}(\ell)$ for the tracers considered in this work. Combined QE + CIB tracers are shown in solid lines, and QE-only tracers in dashed lines. Shaded bands show the $1\sigma$ uncertainty of the per-simulation $A_{\rm lens}^{\rm res}(\ell)$; for clarity, we show them only for the combined QE + CIB cases, but the QE-only cases have comparable uncertainties. Note that these bands are not the spread of the mean, which is smaller by a factor of $1/\sqrt{499}$ and not visible at this scale. Averaging over the multipole range $20 \leq \ell \leq 200$, the standard GMV tracer achieves the lowest $A_{\rm lens}^{\rm res}$ because of its lower QE reconstruction noise, at 47\% combined with CIB, followed by GMVph, then PP. Adding the CIB significantly reduces the $A_{\rm lens}^{\rm res}$ across all cases, by $\approx 20$--$25\%$.
    }
    \label{fig:a_lens_res}
\end{figure}

Since the data auto-spectra agree with the Gaussian simulation means 
(Fig.~\ref{fig:data_LT_specs}), the simulation-derived residual lensing 
amplitudes are expected to be reliable predictors of the data's delensing efficiency.
To quantify the delensing efficiencies of the lensing templates, we compute the resulting residual lensing amplitude, as defined in Eq.~\ref{eq:a_lens_res}.
Fig.~\ref{fig:a_lens_res} shows the scale-dependent $A_{\rm lens}^{\rm res}(\ell)$ for the different tracers considered. As discussed in Sec.~\ref{sec:lensing_template_spectra}, only a fraction of the input lensing $B$-mode power is recovered because tracer noise, masking and apodization, and filtering all suppress or mix modes; these effects are captured in $A_{\rm lens}^{\rm res}$ through their impact on the template's correlation with the input $B$ modes. The standard GMV reconstruction achieves the lowest residual, followed by GMVph and PP, reflecting their relative QE reconstruction noise levels. In all cases, including the CIB significantly improves performance, reducing $A_{\rm lens}^{\rm res}$ by $\approx 20$--$25\%$.

With the fiducial multipole cuts (Sec.~\ref{sec:inputs}), the $A_{\rm lens}^{\rm res}$ are 0.59, 0.61, and 0.69 for the standard GMV, GMVph, and PP tracers, respectively. Including the CIB reduces these to 0.47, 0.48, and 0.52.\footnote{We quote $A_{\rm lens}^{\rm res}$ averaged over $20 \leq \ell \leq 200$. For comparison with the ACT DR6 delensing analysis~\cite{act_lensing_template}, which reports removing 47\% of the lensing $B$-mode power over $30 \leq \ell \leq 300$, our GMVph + CIB template averaged over the same range gives $A_{\rm lens}^{\rm res} = 0.45$ (i.e., 55\% of lensing $B$-mode power removed). However, this is not an apples-to-apples comparison as the ACT value is obtained from delensing observed $B$ modes, whereas ours is a simulation-predicted residual for the lensing template.} The uncertainties on $A_{\rm lens}^{\rm res}$, computed from the spread of simulations, are between 0.021 and 0.025 for all tracers. The GMVph-derived template achieves delensing efficiency comparable to the standard GMV case, particularly when combined with the CIB, while offering improved robustness against foreground-induced biases (Sec.~\ref{sec:foreground_bias}). These values are summarized in Table~\ref{tab:alens_res_fiducial}.

\begin{table}[t]
    \centering
    \caption{Residual $B$-mode lensing amplitude $A_{\rm lens}^{\rm res}$ for the lensing $\phi$ tracers considered in this work. Lower values indicate better delensing efficiency. The values are averaged over 499 Gaussian simulations and over the $\ell$ range $20 \leq \ell \leq 200$, with fiducial multipole cuts imposed.}
    \label{tab:alens_res_fiducial}
    \setlength{\tabcolsep}{9pt}
    \begin{tabular}{lccc}
    \toprule
    QE: & GMV & GMVph & PP \\
    \midrule
    No CIB & 0.59 & 0.61 & 0.69 \\
    + CIB  & 0.47 & 0.48 & 0.52 \\
    \bottomrule
    \end{tabular}
\end{table}

\subsection{Robustness to Analysis Choices} \label{sec:analysis_choices}

In this section, we assess the robustness of our results to variations in  multipole cuts, varying one parameter at a time relative to the fiducial configuration.
The parameters which we vary are: the $\ell_{\rm min}$ and $\ell_{\rm max}$ of the $E$ field, the $L_{\rm min}$ and $L_{\rm max}$ of the CIB tracer, and the $\ell_{\rm max}^{T}$ of the QE reconstruction.

\begin{figure}[t]
	\centering
	\includegraphics[width=\columnwidth]{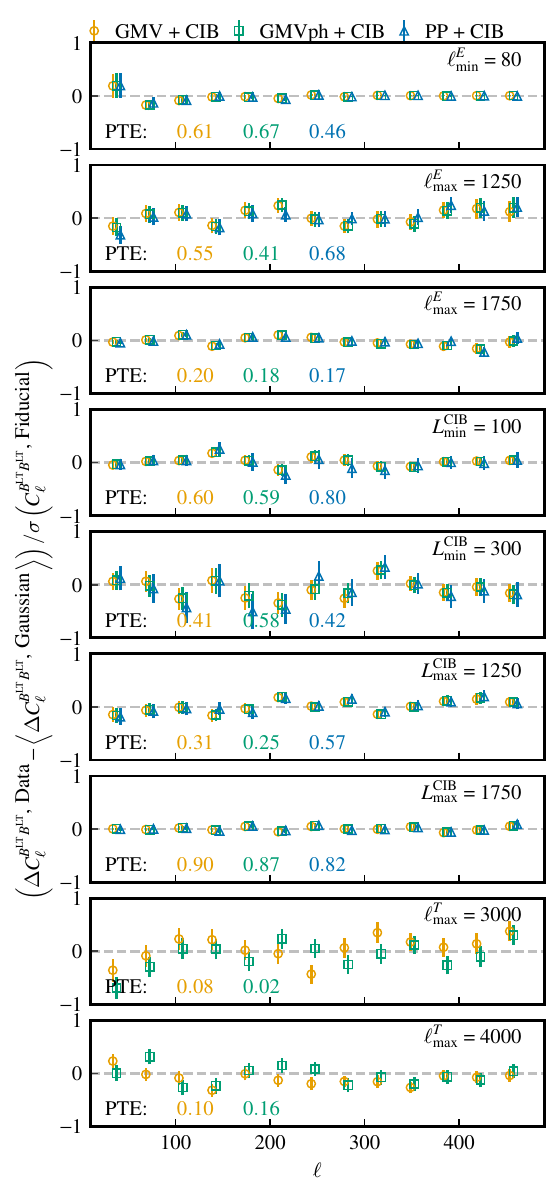}
	\centering \caption{Difference $\Delta C_\ell^{B^\mathrm{LT}B^\mathrm{LT}}$ between the fiducial and modified analysis choices, Gaussian-simulation mean subtracted and normalized by the fiducial uncertainty $\sigma$. For each case, we also report the PTE for data consistency with simulations. Overall, we find that all parameter variations yield results consistent with the baseline.}
	\label{fig:delta_over_sigma}
\end{figure}

Fig.~\ref{fig:delta_over_sigma} shows the difference in lensing template auto-spectra between the fiducial case and each modified case, with the Gaussian-simulation mean subtracted and normalized by the $1\sigma$ uncertainty of the fiducial spectrum. For each case we report the PTE of the data differences with respect to the mean and scatter of the Gaussian simulations, computed as the probability of obtaining a $\chi^2$ at least as large as observed in data, with
\begin{equation}
    \chi^2 = \left(\Delta C_\ell^{\rm data} - \langle \Delta C_\ell^{\rm sim} 
    \rangle\right)^T \mathbb{C}^{-1} \left(\Delta C_\ell^{\rm data} - \langle 
    \Delta C_\ell^{\rm sim} \rangle\right),
\end{equation}
where $\Delta C_\ell = C_\ell^{\rm variant} - C_\ell^{\rm fiducial}$ is the difference in the lensing template auto-spectra between the modified and fiducial cases, and $\mathbb{C}$ is the covariance of $\Delta C_\ell$ estimated from Gaussian simulations.

Increasing the lower bound to $\ell_{\rm min}^{E} = 80$ produces shifts of at most $0.20\sigma$, with a mean shift of $0.04\sigma$ across all tracers,
suggesting that the template is not significantly affected by unmodeled large-scale $E$-mode contamination, such as Galactic foregrounds, in the $\ell$ range between 40 and 80.
We do not run a test with $\ell_{\rm min}^{E}$ decreased because the fiducial value of 40 is near the cutoff scale of the high-pass filtering in the low-$\ell$ maps.
Varying the upper bound to $\ell_{\rm max}^{E} = 1250$ or $1750$ similarly yields no significant shift, with PTEs ranging from 0.17 to 0.68, suggesting the modeling of the transfer function and noise of the low-$\ell$ polarization maps is sufficient.

Varying $L_{\rm min}^{\rm CIB}$ between 100 and 300 and $L_{\rm max}^{\rm CIB}$ between 1250 and 1750 also has little impact on the template. The PTEs range from 0.25 to 0.90 across all tracers, reflecting good agreement between data differences with simulations as described in Sec.~\ref{sec:cib_sims}. The $L_{\rm min}^{\rm CIB}$ test additionally indicates that residual Galactic foregrounds in the CIB map are subdominant~\cite[e.g.][]{anton_delens_cib_dust}.

Finally, we vary the temperature multipole cutoff $\ell_{\rm max}^{T}$ used in the QE reconstruction as a probe of the impact of unmodeled non-Gaussian extragalactic foreground biases. For both the GMV + CIB and GMVph + CIB tracers, the PTEs at $\ell_{\rm max}^{T} = 3000$ and $4000$ do not indicate a statistically significant difference from the fiducial configuration, implying that any foreground bias introduced by the additional temperature modes in the fiducial case is not detectable above the level of statistical fluctuations in our Gaussian simulations.
For standard GMV (no foreground mitigation), while the PTE indicates that there is no statistically significant difference between the difference in simulations and data when going to $\ell_{\rm max}^{T} = 4000$, the points are consistently below 0, which indicates a low level of bias.
Because the $\chi^2$ statistic is insensitive to coherent one-sided deviations, we also compute a PTE derived from a $\chi$ test.
For standard GMV at $\ell_{\rm max}^{T} = 4000$, the $\chi$ PTE is 0.02, lower than the 0.41 for GMVph, which may reflect a mild coherent offset compared to simulation differences in the standard GMV case, though we do not interpret this as a statistically significant bias (for $\ell_{\rm max}^{T} = 3000$, the $\chi$ PTEs are 0.42 and 0.26, respectively, for GMV and GMVph).
We further discuss the effects of extragalactic foreground bias in the next section.

We conclude that the delensing performance is stable under reasonable variations of analysis choices, particularly for the GMVph + CIB tracer.

\subsection{Foreground Bias} \label{sec:foreground_bias}

We next assess the impact of non-Gaussian foregrounds on the lensing templates using \agora\ simulations, which include extragalactic foregrounds such as the tSZ/kSZ effects, CIB, and radio sources, which are realistically correlated with the large-scale structure responsible for CMB lensing.
These foregrounds introduce biases in the reconstructed lensing potential $\hat{\phi}$ through higher-order correlations, which then propagate into the lensing templates, biasing lensing template power spectra and modifying the correlation with the true lensing $B$-mode signal \cite{baleato_lizancos_2022}.

\begin{figure}[t]
	\centering
	\includegraphics[width=\columnwidth]{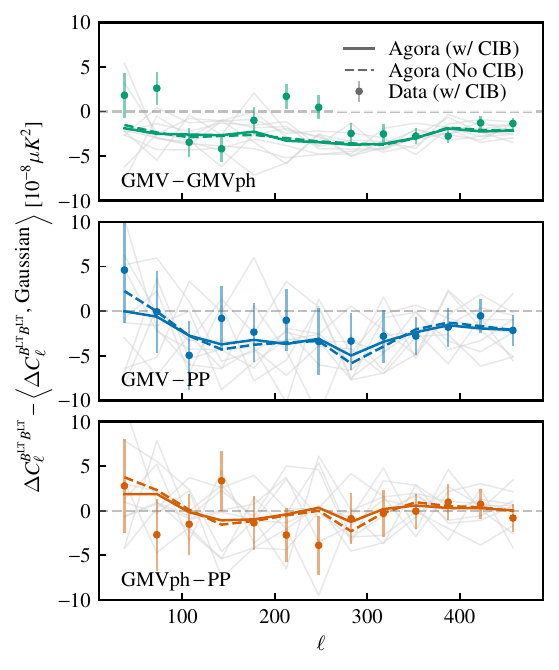}
	\caption{Gaussian-simulation mean subtracted differences in lensing template auto-spectra for \agora\ simulations (averaged over 10 realizations) and data. The \agora\ differences are presented in solid~(with CIB) and dashed~(without CIB) lines, and the data points~(with CIB) are plotted with error bars showing measurement uncertainty. The light grey lines show the per-realization \agora\ differences for the with-CIB case.
    Differences between the standard GMV and GMVph~(top) and between the standard GMV and PP~(middle) show clear offsets from zero, consistent with residual foreground bias in the standard GMV $\phi$ reconstruction. The differences between GMVph and PP reconstructions~(bottom), on the other hand, are consistent with zero. Since the PP estimator is expected to have negligible foreground-induced bias, this consistency suggests that the GMVph reconstruction also has insignificant foreground bias.}
	\label{fig:agora_data_spec_diff}
\end{figure}

We begin by comparing relative differences between \agora\ lensing template spectra constructed using different tracers. In Fig.~\ref{fig:agora_data_spec_diff}, the lines show the differences between lensing template auto-spectra for the GMV, GMVph, and PP reconstructions as the $\phi$ tracers, averaged over 10 \agora\ patches. The quantities shown are Gaussian-simulation mean subtracted to isolate deviations from the Gaussian expectation:
\begin{equation} \label{eq:diff_agora_mean_sub}
    \langle \Delta C_\ell^{BB,\ \rm Agora} \rangle - \langle \Delta C_\ell^{BB,\ \rm Gaussian} \rangle,
\end{equation}
where the $\Delta$s denote differences between estimator choices. The points with error bars show the same quantities for data.

Differences involving the GMV reconstruction (top and middle panels of Fig.~\ref{fig:agora_data_spec_diff}) show clear deviations from zero, consistent with foreground-induced bias in the $\hat{\phi}$ reconstruction. In contrast, the GMVph and PP reconstructions (bottom panel) are consistent with each other.
Quantitatively, for the 10-\agora-patch average (with the covariance scaled accordingly by $1/10$), the PTEs are $\approx 10^{-87}$, $\approx 10^{-8}$, and 0.79 for the GMV$-$GMVph, GMV$-$PP, and GMVph$-$PP differences, respectively, indicating that the former two are inconsistent with Gaussian expectations, while the latter is consistent.\footnote{The ten \agora\ patches are not independent, since the \agora\ lightcones are constructed by tiling simulation boxes, thus structures can repeat across the sky. This means the $1/10$ scaling of the covariance can underestimate the the 10-patch-mean scatter and thus overstate the significance of the averaged-patch PTEs; the per-patch PTEs, which do not involve this scaling, are less extreme but lead to the same conclusions.} In data, the corresponding PTEs are $\approx 10^{-5}$, 0.91, and 0.97, with the GMVph$-$PP case remaining consistent with Gaussian expectations and the GMV$-$GMVph case showing a deviation.

We next check the $\chi$ PTE to better isolate one-sided differences.
The $\chi$ PTE, which is sensitive to coherent one-sided deviations that the $\chi^2$ statistic may miss, more clearly captures the potential data difference sourced by foregrounds: for the 10-\agora-patch average, the $\chi$ PTEs are $< 0.002$, $< 0.002$, and 0.78 for the GMV$-$GMVph, GMV$-$PP, and GMVph$-$PP cases, respectively,\footnote{In contrast to the $\chi^2$-based PTE, which is evaluated analytically from the $\chi^2$ distribution and is therefore not limited in resolution, the $\chi$ PTE is estimated empirically as the fraction of Gaussian simulations whose summed normalized residuals (in absolute value) meet or exceed that of data. With 499 Gaussian simulations, the statistic can only take discrete values, so we report a $\chi$ PTE of 0 as PTE $< 1/499 \approx 0.002$.} while in data they are 0.002, 0.12, and 0.66. The $\chi$ PTE indicates that both the GMV$-$PP and GMVph$-$PP differences in data are in good agreement with zero.
Since the PP estimator is expected to be largely insensitive to extragalactic foregrounds, this provides evidence that neither the GMV + CIB nor GMVph + CIB constructed lensing templates have detectable levels of bias given the size of the difference error bars.
However, in the GMV$-$GMVph case, the PTE indicates a coherent offset in the lensing template bandpowers between the two. This means that either GMV or GMVph are less consistent with PP. As indicated by the lower $\chi$ PTE value and that the GMVph$-$PP points lie more consistently around zero (bottom panel of Fig.~\ref{fig:agora_data_spec_diff}), this suggests that there is a low level of bias in the GMV case that is insignificant given the larger difference error bars when PP is differenced.

We end the discussion of lensing template difference tests by comparing the size of the difference between data and \agora\ simulations for cases including GMV to gain a sense of the relative size of bias in data vs. that in \agora\ simulations. For the GMV$-$GMVph and GMV$-$PP cases, the absolute magnitudes of the data differences are comparable to their \agora\ counterparts given statistical uncertainties, with many of the bins showing smaller differences. This suggests that the level of foreground bias in the data is at most what is seen in the \agora\ simulations and likely smaller.

While these comparisons establish relative behavior between estimator choices, they do not isolate the effect of foregrounds alone. Differences between reconstruction methods also reflect variations in reconstruction noise, filtering, and weighting, making it less directly interpretable as a measure of foreground bias.
To better isolate the foreground contribution, we compare fiducial \agora\ simulations with a matched set in which the foregrounds are replaced by Gaussian realizations with identical power.
These Gaussian foregrounds lack the higher-order statistics that source lensing reconstruction bias, and therefore do not bias the lensing template.
Since both simulation sets share the same realizations of the lensed CMB and instrumental noise, differences can be directly attributed to foreground non-Gaussianity.

\begin{figure}[t]
	\centering
	\includegraphics[width=\columnwidth]{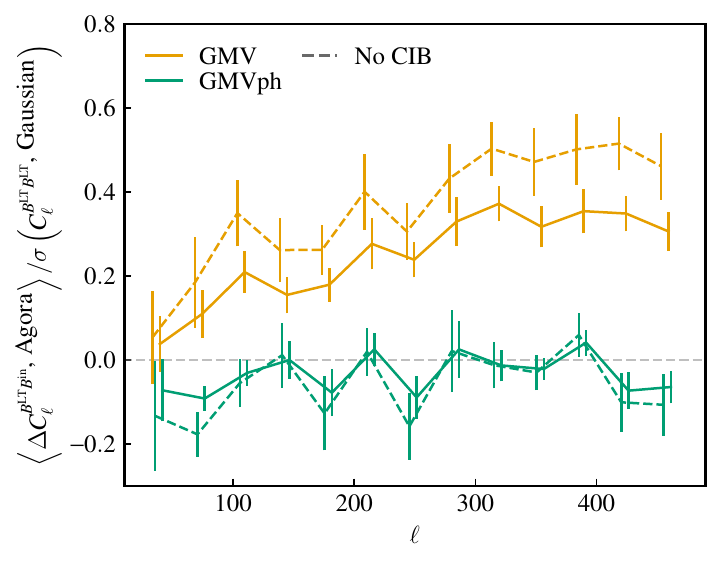}
	\caption{Foreground-induced bias in the lensing template, computed as the difference in the template cross-spectrum with the input $B$, between fiducial and Gaussian-foreground \agora\ simulations. Here we normalize by the standard deviation of the template auto-spectrum, as seen also in the error bars of Fig.~\ref{fig:data_LT_specs}. Dashed lines show QE-only cases; solid lines include the CIB tracer. We average over 10 \agora\ patches, and show standard error (i.e., divided by $\sqrt{N}$) in the error bars.
    The GMVph case reduces bias from foreground non-Gaussianity. Combined with CIB, the foreground-induced bias becomes $\lesssim 0.1\sigma$ with an average of $\approx 0.05\sigma$.
    }
	\label{fig:agora_fg_bias_ratio_sigma}
\end{figure}

Fig.~\ref{fig:agora_fg_bias_ratio_sigma} shows the resulting difference in the lensing template cross-spectrum with the input $B$ field, normalized by the standard deviation of the template auto-spectrum from Gaussian simulations. This provides an absolute measure of the foreground-induced bias in simulations in units of the statistical uncertainty.

We find that the lensing template constructed with GMV exhibits a clear bias from foreground non-Gaussianity, particularly at small angular scales, consistent with contamination in the temperature maps. Using GMVph significantly reduces this bias, and the inclusion of the CIB further suppresses it by reducing the relative contribution of the biased CMB-based reconstruction.
This is a clear demonstration of the effectiveness of the hardening procedure in mitigating foreground-induced mode coupling.

For the GMVph + CIB case, the foreground-induced bias is $\lesssim 0.1\sigma$ across the relevant multipole range, with an average level of $\approx 0.05\sigma$. Since the data show deviations smaller than or comparable to those in \agora\ (Fig.~\ref{fig:agora_data_spec_diff}), this indicates that the foreground-induced bias in the data is at most at this level and likely smaller.

Overall, these results demonstrate that foreground-induced biases in the lensing templates are subdominant compared to statistical uncertainties for the GMVph + CIB combined tracer, supporting its use as a robust input for delensing.

\subsection{Impact on $\sigma(r)$}

We now connect the performance of the lensing templates to their impact on constraints on the tensor-to-scalar ratio $r$.
The statistical uncertainty on $r$ is set in part by the residual lensing $B$-mode power after delensing. Reducing $A_{\rm lens}^{\rm res}$ lowers the lensing contribution to the $B$-mode variance and thus directly improves $\sigma(r)$, with the level of improvement depending on the relative contributions of lensing, instrumental noise, and foreground residuals.

As an example, we lay out a possible improvement in $\sigma(r)$ for the BK18 dataset~\cite{bk18} if delensed with the lensing template constructed using the baseline GMVph + CIB tracer, with $A_{\rm lens}^{\rm res} \simeq 0.48$.
For the BK18 dataset, the total $\sigma(r)$ is 0.009. When the same analysis is run on simulations with no lensing $B$ modes, the $\sigma(r)$ becomes 0.004. Therefore, lensing contributes $\sigma(r) \approx 0.005$ of the total uncertainty of $r$. Assuming that the $A_{\rm lens}^{\rm res}$ measured in this work against input $B$ modes are accurate representations of the mode overlap with the BICEP-measured modes,\footnote{This is a toy scenario to provide a back-of-the-envelope estimate. In practice, there will be lensing modes in the BICEP maps that are not reconstructed by this lensing template.} then we expect to reduce the lensing $\sigma(r)$ to 0.0024, leading to a total $\sigma(r)$ of 0.0064 for a BK18-delensed analysis using the GMVph + CIB template (a roughly $30\%$ improvement). Looking to the future, the next BICEP $r$ result's $\sigma(r)$ will be even more lensing dominated, so delensing using this same template will lead to a larger fractional improvement on $\sigma(r)$.

\section{Conclusion} \label{sec:conclusion}

In this work, we report the construction and validation of a foreground-robust CMB lensing $B$-mode template with the highest delensing efficiency to date, with a simulation-predicted residual lensing amplitude of $A_{\rm lens}^{\rm res} \simeq 0.48$, made possible by the high correlation of the SPT-3G D1 lensing reconstruction with the true lensing field, together with the complementary CIB $\phi$ tracer.
We construct templates and perform tests using SPT-3G D1 data for $E$ modes and CMB-reconstructed $\phi$ using GMV, GMVph, and PP estimators, with and without the inclusion of a \planck\ CIB $\phi$ tracer.

We find that the lensing template data auto-spectra agree with Gaussian simulation expectations, confirming that the signal and noise modeling used throughout this work accurately describes the data, and thus that the simulation-derived filtering and weighting are also near-optimal for data. Combining internal CMB lensing reconstructions with the CIB exploits their complementary scale dependence in $\rho^{\phi}_L$, resulting in $A_{\rm lens}^{\rm res} \simeq 0.48$ for the GMVph + CIB tracer, which corresponds to a $\approx 21\%$ improvement over the QE-only tracer. These results are robust to reasonable variations in analysis choices.

We use \agora\ simulations and lensing template difference-spectra to assess the impact of non-Gaussian extragalactic foregrounds.
We find that foreground-induced biases in the lensing template are significant for the standard GMV-constructed templates, but are strongly suppressed for the GMVph and PP reconstructions. For the GMVph + CIB constructed template, the residual bias is $\lesssim 0.1$ of the template bandpower statistical uncertainty, and is comparable or smaller in data than in the simulations, indicating that foreground contamination is well controlled.

The achieved level of delensing directly reduces the lensing contribution to the $B$-mode variance, leading to a corresponding improvement in $\sigma(r)$ in regimes where lensing sample variance is important. At the same time, the foreground-induced bias in the lensing template is subdominant compared to statistical uncertainties, and therefore does not represent a limiting systematic for current and near-term analyses.

As primordial $B$-mode searches become increasingly lensing-dominated, delensing will be essential to further improving sensitivity to $r$~\cite{belkner2024,aso_r}. The lensing template demonstrated here, combining foreground-robust CMB lensing reconstructions with an external CIB tracer, achieves a level of delensing that will substantially improve $\sigma(r)$ for the upcoming BICEP dataset, where lensing contributes more than half of the total variance. 
Looking forward, delensing is crucial for the South Pole Observatory (South Pole Observatory et al., {\it in prep.}) to reach its goal of $\sigma(r) \lesssim$ 0.001; a foreground-robust, high-efficiency lensing template is a step towards this goal.

\appendix

\section{Filtering and Tracer Weighting} \label{sec:filtering_and_weighting}

In this appendix, we first show that the lensing template auto- and cross-spectra based on perturbative lensing are identical in the idealized case where the signal and noise for $E$ and $\phi$ are statistically isotropic and when the Wiener filter correctly describes both terms.
Then we show that in our realistic simulations, where statistical isotropy is broken from masking and filtering, the lensing template auto- and cross-spectra are within 3\% of each other, demonstrating that the Wiener filters applied to the $\phi$ tracers and $E$ modes are near-optimal.
This also substantiates the use of the correlation coefficient $\rho_{\ell}^B$ in reporting $A_{\rm lens}^{\rm res}$, our metric for delensing efficiency.

We start from Eq.~\ref{eq:B_lens} for the input $B$ and Eq.~\ref{eq:B_template} for the template $B$ and write the cross-spectrum as
\begin{equation} \label{eq:clbb_cross}
    C_\ell^{B^{\rm LT}B^{\rm in}} \propto \sum_{\ell' L} (g^{EB}_{\ell \ell' L})^2 \langle E^{\rm WF}_{\ell' m'} E^{*}_{\ell' m'} \rangle \langle \phi^{\rm WF}_{LM} \phi^{*}_{LM} \rangle
\end{equation}
and the auto-spectrum as
\begin{equation} \label{eq:clbb_auto}
    C_\ell^{B^{\rm LT}B^{\rm LT}} \propto \sum_{\ell' L} (g^{EB}_{\ell \ell' L})^2 \langle E^{\rm WF}_{\ell' m'} E^{\rm WF*}_{\ell' m'} \rangle \langle \phi^{\rm WF}_{LM} \phi^{\rm WF *}_{LM} \rangle,
\end{equation}
where we have assumed uncorrelated $E$ and $\phi$, i.e., $\langle E_{\ell m} \phi^{*}_{LM} \rangle = 0$.
We then explicitly write the Wiener-filtered tracer as
\begin{equation} \label{eq:phi_wf}
    \phi^{\rm WF}_{LM} = W^{\phi}_L (\phi_{LM} + n^{\phi}_{LM})
\end{equation}
where $\phi$, $n^{\phi}$ are the signal and noise parts of the tracer, and the $\phi$ Wiener filter $W^{\phi}_L$ for QE is defined as in Eq.~\ref{eq:qe_wiener_filter}.
Similarly for the Wiener-filtered $E$,
\begin{equation} \label{eq:e_wf}
    E^{\rm WF}_{\ell m} = W^{E}_\ell (E_{\ell m} + n^{E}_{\ell m}).
\end{equation}
Then, assuming that the noise is uncorrelated with the true $E$ and $\phi$, Eq.~\ref{eq:clbb_cross} becomes
\begin{equation}
    C_\ell^{B^{\rm LT}B^{\rm in}} \propto \sum_{\ell' L} (g^{EB}_{\ell \ell' L})^2  W^{E}_{\ell'} C_{\ell'}^{EE} W^{\phi}_{L} C_{L}^{\phi\phi},
\end{equation}
and for Eq.~\ref{eq:clbb_auto} we have
\begin{multline}
    C_\ell^{B^{\rm LT}B^{\rm LT}} \propto \sum_{\ell' L} (g^{EB}_{\ell \ell' L})^2 (W^{E}_{\ell'})^2 (C_{\ell'}^{EE}+N_{\ell'}^{EE})\\
    \times (W^{\phi}_{L})^2 (C_{L}^{\phi\phi}+N_{L}^{\phi\phi})
\end{multline}
Using the definitions of the Wiener filters $W^{\phi}_L$ and $W^{E}_\ell$, this reduces to
\begin{equation}
    C_\ell^{B^{\rm LT}B^{\rm LT}} \propto \sum_{\ell' L} (g^{EB}_{\ell \ell' L})^2  W^{E}_{\ell'} C_{\ell'}^{EE} W^{\phi}_{L} C_{L}^{\phi\phi};
\end{equation}
thus, when the filtering is optimal, $C_\ell^{B^{\rm LT}B^{\rm in}} = C_\ell^{B^{\rm LT}B^{\rm LT}}$.\footnote{We note that the equality also holds for $E_{\ell m}$ and $\phi_{L M}$ that have isotropic biases, as long as the signal variance in their respective filters correctly describe them.}

\begin{figure}[t]
	\centering
	\includegraphics[width=\columnwidth]{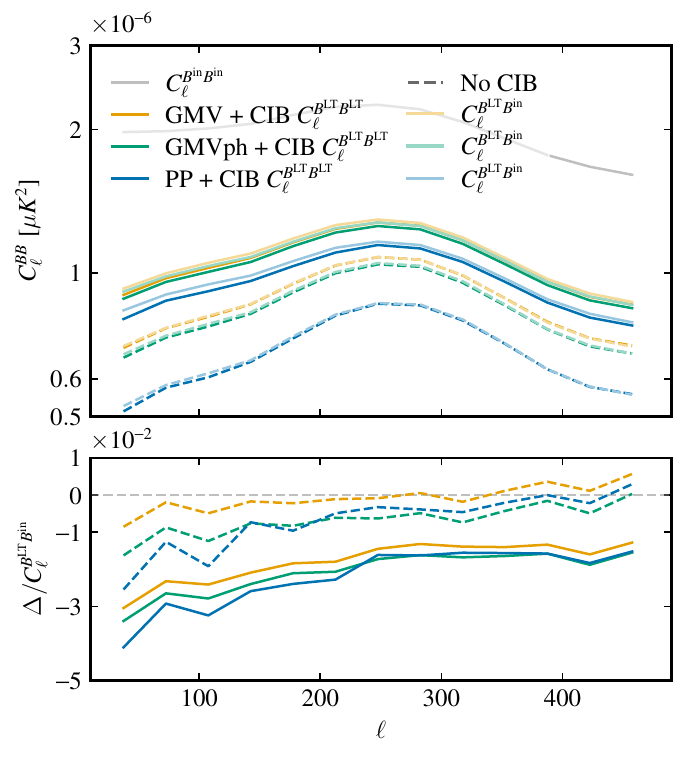}
	\centering \caption{The power spectra of the lensing templates from Gaussian simulations, constructed using the standard GMV, GMVph, and PP as the $\phi$ tracer. We include cases with (solid lines) and without (dashed lines) the CIB tracer. The lensing template auto-spectra are shown in darker colors, and the cross-spectra between the template and the input $B$ are shown in lighter colors. The gray curve corresponds to the input $B$ auto-spectrum. All spectra are averaged over 499 simulations. The bottom subplot shows the fractional difference between the auto- and cross-spectra for each case. The auto- and cross-spectra match each other within $\approx 3\%$.
    }
	\label{fig:sims_LT_specs}
\end{figure}

Fig.~\ref{fig:sims_LT_specs} shows the lensing template power spectra for the different $\phi$ tracers considered in this work, computed from 499 Gaussian simulations.
Since the input lensing signal is known for each Gaussian simulation, we can directly compute the cross-spectra between the constructed lensing templates and the input $B$-mode field. This allows us to verify that the template construction is implemented correctly and that the filtering and weighting behave as expected.

The auto- and cross-spectra of the simulations agree to within $\approx 3\%$ across all scales shown and across all tracer choices, providing confirmation that the filtering and weighting used in the template construction are near-optimal.
To check that the remaining difference is sourced primarily by mode mismatch at the apodization area of the mask, we performed the auto-cross comparison using a modified mask when computing the spectra. We kept the same set of interior point source holes, but reduced the outer border of the mask further inward. With this smaller mask, the auto- and cross-spectra agree more closely than in our fiducial case.
This is consistent with the interpretation that the $\mathbb{C}^{-1}$ filtering applied on the apodized mask downweights modes near the boundary, and this produces a mismatch between $B^{\rm LT}$ and $B^{\rm in}$ (which has not gone through $\mathbb{C}^{-1}$ filtering from $Q/U$ but is directly masked in $B$ modes). By using a smaller mask for the spectrum computation, we exclude the boundary region where this mismatch is the largest, bringing the auto- and cross-spectra into better agreement. Thus this confirms that the residual disagreement in the fiducial case is driven by boundary effects from the mask apodization, rather than a miscalibration of the filters.
\\

\acknowledgments
We thank Clem Pryke, Colin Bischoff, John Kovac, Zeeshan Ahmed, Matthew Petroff, and Jamie Cheshire for valuable comments to a draft of this paper, and members of the BICEP collaboration for useful discussions throughout this project.
Much of the computing for this project was performed on the Sherlock cluster. We would like to thank Stanford University and the Stanford Research Computing Center for providing computational resources and support that contributed to these research results.
We also acknowledge the computing resources provided on Crossover, a high-performance computing cluster operated by the Laboratory Computing Resource Center at Argonne National Laboratory.
WLKW acknowledges support from an Early Career Research Award DE-SC0026376 of the Department of Energy.
The South Pole Telescope program is supported by the National Science Foundation (NSF) through awards OPP-1852617 and OPP-2332483. Partial support is also provided by the Kavli Institute of Cosmological Physics at the University of Chicago. 
Argonne National Laboratory’s work was supported by the U.S. Department of Energy, Office of High Energy Physics, under contract DE-AC02-06CH11357. 
The UC Davis group acknowledges support from Michael and Ester Vaida. 
Work at the Fermi National Accelerator Laboratory (Fermilab), a U.S. Department of Energy, Office of Science, Office of High Energy Physics HEP User Facility, is managed by Fermi Forward Discovery Group, LLC, acting under Contract No. 89243024CSC000002.
The Melbourne authors acknowledge support from the Australian Research Council’s Discovery Project scheme (No. DP260100705). 
The Paris group has received funding from the European Research Council (ERC) under the European Union’s Horizon 2020 research and innovation program (grant agreement No. 101001897), and funding from the Centre National d’Etudes Spatiales. 
The SLAC group is supported in part by the Department of Energy at SLAC National Accelerator Laboratory, under contract DE-AC02-76SF00515.

Some of the results in this paper have been derived using the healpy and HEALPix package.

\bibliography{biblio.bib}

\end{document}

%% file: authors.tex
\affiliation{Department of Physics, Stanford University, 382 Via Pueblo Mall, Stanford, CA, 94305, USA}
\affiliation{California Institute of Technology, 1200 East California Boulevard., Pasadena, CA, 91125, USA}
\affiliation{Kavli Institute for Particle Astrophysics and Cosmology, Stanford University, 452 Lomita Mall, Stanford, CA, 94305, USA}
\affiliation{SLAC National Accelerator Laboratory, 2575 Sand Hill Road, Menlo Park, CA, 94025, USA}
\affiliation{Department of Astronomy and Astrophysics, University of Chicago, 5640 South Ellis Avenue, Chicago, IL, 60637, USA}
\affiliation{Kavli Institute for Cosmological Physics, University of Chicago, 5640 South Ellis Avenue, Chicago, IL, 60637, USA}
\affiliation{NSF-Simons AI Institute for the Sky (SkAI), 172 E. Chestnut St., Chicago, IL 60611, USA}
\affiliation{Department of Statistics, University of California, One Shields Avenue, Davis, CA 95616, USA}
\affiliation{Fermi National Accelerator Laboratory, MS209, P.O. Box 500, Batavia, IL, 60510, USA}
\affiliation{School of Physics, University of Melbourne, Parkville, VIC 3010, Australia}
\affiliation{Sorbonne Universit\'e, CNRS, UMR 7095, Institut d'Astrophysique de Paris, 98 bis bd Arago, 75014 Paris, France}
\affiliation{Department of Physics and Astronomy, University of New Mexico, Albuquerque, NM, 87131, USA}
\affiliation{High-Energy Physics Division, Argonne National Laboratory, 9700 South Cass Avenue, Lemont, IL, 60439, USA}
\affiliation{University Observatory, Faculty of Physics, LMU Munich, Scheinerstr.~1, 81679 Munich, Germany}
\affiliation{Enrico Fermi Institute, University of Chicago, 5640 South Ellis Avenue, Chicago, IL, 60637, USA}
\affiliation{Department of Physics, University of Chicago, 5640 South Ellis Avenue, Chicago, IL, 60637, USA}
\affiliation{Istituto ricerche solari Aldo e Cele Dacc\`o (IRSOL), Faculty of Informatics, Universit\`a della Svizzera italiana, 6605 Locarno, Switzerland}
\affiliation{Universit\'e de Gen\`eve, D\'epartement de Physique Th\'eorique, 24 Quai Ansermet, CH-1211 Gen\`eve 4, Switzerland}
\affiliation{National Taiwan University, No. 1, Sec. 4, Roosevelt Road, Taipei 106319, Taiwan}
\affiliation{Department of Physics, University of California, Berkeley, CA, 94720, USA}
\affiliation{Universit\'e Paris-Saclay, Universit\'e Paris Cit\'e, CEA, CNRS, AIM, 91191, Gif-sur-Yvette, France}
\affiliation{Department of Astronomy, University of Illinois Urbana-Champaign, 1002 West Green Street, Urbana, IL, 61801, USA}
\affiliation{High Energy Accelerator Research Organization (KEK), Tsukuba, Ibaraki 305-0801, Japan}
\affiliation{Department of Physics and McGill Space Institute, McGill University, 3600 Rue University, Montreal, Quebec H3A 2T8, Canada}
\affiliation{Canadian Institute for Advanced Research, CIFAR Program in Gravity and the Extreme Universe, Toronto, ON, M5G 1Z8, Canada}
\affiliation{Joseph Henry Laboratories of Physics, Jadwin Hall, Princeton University, Princeton, NJ 08544, USA}
\affiliation{Department of Astronomy, University of Science and Technology of China, Hefei 230026, China}
\affiliation{School of Astronomy and Space Science, University of Science and Technology of China, Hefei 230026}
\affiliation{Department of Physics, University of Illinois Urbana-Champaign, 1110 West Green Street, Urbana, IL, 61801, USA}
\affiliation{Department of Physics and Astronomy, University of California, Los Angeles, CA, 90095, USA}
\affiliation{Department of Physics and Astronomy, Michigan State University, East Lansing, MI 48824, USA}
\affiliation{Department of Physics \& Astronomy, University of California, One Shields Avenue, Davis, CA 95616, USA}
\affiliation{CASA, Department of Astrophysical and Planetary Sciences, University of Colorado, Boulder, CO, 80309, USA }
\affiliation{Department of Physics, University of Colorado, Boulder, CO, 80309, USA}
\affiliation{Department of Physics \& Astronomy, Box 41051, Texas Tech University, Lubbock TX 79409-1051, USA}
\affiliation{Center for AstroPhysical Surveys, National Center for Supercomputing Applications, Urbana, IL, 61801, USA}
\affiliation{Department of Physics, Case Western Reserve University, Cleveland, OH, 44106, USA}
\affiliation{Department of Physics, Villanova University, 800 E Lancaster Ave, Villanova, PA 19085, USA}
\affiliation{Center for Astrophysics \textbar{} Harvard \& Smithsonian, 60 Garden Street, Cambridge, MA, 02138, USA}
\affiliation{CERCA/ISO, Department of Physics, Case Western Reserve University, Cleveland, OH 44106, USA}
\affiliation{School of Physics and Astronomy, Cardiff University, Cardiff, CF24 3AA, UK}
\author{Y.~Nakato}
\email{yukanaka@stanford.edu}
\affiliation{Department of Physics, Stanford University, 382 Via Pueblo Mall, Stanford, CA, 94305, USA}
\author{W.~L.~K.~Wu\,\orcidlink{0000-0001-5411-6920}}
\affiliation{California Institute of Technology, 1200 East California Boulevard., Pasadena, CA, 91125, USA}
\affiliation{Kavli Institute for Particle Astrophysics and Cosmology, Stanford University, 452 Lomita Mall, Stanford, CA, 94305, USA}
\affiliation{SLAC National Accelerator Laboratory, 2575 Sand Hill Road, Menlo Park, CA, 94025, USA}
\author{Y.~Omori\,\orcidlink{0000-0002-0963-7310}}
\affiliation{Department of Astronomy and Astrophysics, University of Chicago, 5640 South Ellis Avenue, Chicago, IL, 60637, USA}
\affiliation{Kavli Institute for Cosmological Physics, University of Chicago, 5640 South Ellis Avenue, Chicago, IL, 60637, USA}
\affiliation{NSF-Simons AI Institute for the Sky (SkAI), 172 E. Chestnut St., Chicago, IL 60611, USA}
\author{E.~Anderes\,\orcidlink{0009-0003-3245-3979}}
\affiliation{Department of Statistics, University of California, One Shields Avenue, Davis, CA 95616, USA}
\author{A.~J.~Anderson\,\orcidlink{0000-0002-4435-4623}}
\affiliation{Fermi National Accelerator Laboratory, MS209, P.O. Box 500, Batavia, IL, 60510, USA}
\affiliation{Kavli Institute for Cosmological Physics, University of Chicago, 5640 South Ellis Avenue, Chicago, IL, 60637, USA}
\affiliation{Department of Astronomy and Astrophysics, University of Chicago, 5640 South Ellis Avenue, Chicago, IL, 60637, USA}
\author{B.~Ansarinejad}
\affiliation{School of Physics, University of Melbourne, Parkville, VIC 3010, Australia}
\author{M.~Archipley\,\orcidlink{0000-0002-0517-9842}}
\affiliation{Department of Astronomy and Astrophysics, University of Chicago, 5640 South Ellis Avenue, Chicago, IL, 60637, USA}
\affiliation{Kavli Institute for Cosmological Physics, University of Chicago, 5640 South Ellis Avenue, Chicago, IL, 60637, USA}
\author{L.~Balkenhol\,\orcidlink{0000-0001-6899-1873}}
\affiliation{Sorbonne Universit\'e, CNRS, UMR 7095, Institut d'Astrophysique de Paris, 98 bis bd Arago, 75014 Paris, France}
\author{D.~R.~Barron\,\orcidlink{0000-0002-1623-5651}}
\affiliation{Department of Physics and Astronomy, University of New Mexico, Albuquerque, NM, 87131, USA}
\author{P.~S.~Barry\,\orcidlink{0000-0001-9103-9354}}
\affiliation{School of Physics and Astronomy, Cardiff University, Cardiff, CF24 3AA, UK}
\author{K.~Benabed}
\affiliation{Sorbonne Universit\'e, CNRS, UMR 7095, Institut d'Astrophysique de Paris, 98 bis bd Arago, 75014 Paris, France}
\author{A.~N.~Bender\,\orcidlink{0000-0001-5868-0748}}
\affiliation{High-Energy Physics Division, Argonne National Laboratory, 9700 South Cass Avenue, Lemont, IL, 60439, USA}
\affiliation{Kavli Institute for Cosmological Physics, University of Chicago, 5640 South Ellis Avenue, Chicago, IL, 60637, USA}
\affiliation{Department of Astronomy and Astrophysics, University of Chicago, 5640 South Ellis Avenue, Chicago, IL, 60637, USA}
\author{B.~A.~Benson\,\orcidlink{0000-0002-5108-6823}}
\affiliation{Fermi National Accelerator Laboratory, MS209, P.O. Box 500, Batavia, IL, 60510, USA}
\affiliation{Kavli Institute for Cosmological Physics, University of Chicago, 5640 South Ellis Avenue, Chicago, IL, 60637, USA}
\affiliation{Department of Astronomy and Astrophysics, University of Chicago, 5640 South Ellis Avenue, Chicago, IL, 60637, USA}
\author{F.~Bianchini\,\orcidlink{0000-0003-4847-3483}}
\affiliation{Kavli Institute for Particle Astrophysics and Cosmology, Stanford University, 452 Lomita Mall, Stanford, CA, 94305, USA}
\affiliation{Department of Physics, Stanford University, 382 Via Pueblo Mall, Stanford, CA, 94305, USA}
\affiliation{SLAC National Accelerator Laboratory, 2575 Sand Hill Road, Menlo Park, CA, 94025, USA}
\author{L.~E.~Bleem\,\orcidlink{0000-0001-7665-5079}}
\affiliation{High-Energy Physics Division, Argonne National Laboratory, 9700 South Cass Avenue, Lemont, IL, 60439, USA}
\affiliation{Kavli Institute for Cosmological Physics, University of Chicago, 5640 South Ellis Avenue, Chicago, IL, 60637, USA}
\affiliation{Department of Astronomy and Astrophysics, University of Chicago, 5640 South Ellis Avenue, Chicago, IL, 60637, USA}
\author{S.~Bocquet\,\orcidlink{0000-0002-4900-805X}}
\affiliation{University Observatory, Faculty of Physics, LMU Munich, Scheinerstr.~1, 81679 Munich, Germany}
\author{F.~R.~Bouchet\,\orcidlink{0000-0002-8051-2924}}
\affiliation{Sorbonne Universit\'e, CNRS, UMR 7095, Institut d'Astrophysique de Paris, 98 bis bd Arago, 75014 Paris, France}
\author{E.~Camphuis\,\orcidlink{0000-0003-3483-8461}}
\affiliation{Sorbonne Universit\'e, CNRS, UMR 7095, Institut d'Astrophysique de Paris, 98 bis bd Arago, 75014 Paris, France}
\author{M.~G.~Campitiello}
\affiliation{High-Energy Physics Division, Argonne National Laboratory, 9700 South Cass Avenue, Lemont, IL, 60439, USA}
\author{J.~E.~Carlstrom\,\orcidlink{0000-0002-2044-7665}}
\affiliation{Kavli Institute for Cosmological Physics, University of Chicago, 5640 South Ellis Avenue, Chicago, IL, 60637, USA}
\affiliation{Enrico Fermi Institute, University of Chicago, 5640 South Ellis Avenue, Chicago, IL, 60637, USA}
\affiliation{Department of Physics, University of Chicago, 5640 South Ellis Avenue, Chicago, IL, 60637, USA}
\affiliation{High-Energy Physics Division, Argonne National Laboratory, 9700 South Cass Avenue, Lemont, IL, 60439, USA}
\affiliation{Department of Astronomy and Astrophysics, University of Chicago, 5640 South Ellis Avenue, Chicago, IL, 60637, USA}
\author{J.~Carron\,\orcidlink{0000-0002-5751-1392}}
\affiliation{Istituto ricerche solari Aldo e Cele Dacc\`o (IRSOL), Faculty of Informatics, Universit\`a della Svizzera italiana, 6605 Locarno, Switzerland}
\affiliation{Universit\'e de Gen\`eve, D\'epartement de Physique Th\'eorique, 24 Quai Ansermet, CH-1211 Gen\`eve 4, Switzerland}
\author{C.~L.~Chang}
\affiliation{High-Energy Physics Division, Argonne National Laboratory, 9700 South Cass Avenue, Lemont, IL, 60439, USA}
\affiliation{Kavli Institute for Cosmological Physics, University of Chicago, 5640 South Ellis Avenue, Chicago, IL, 60637, USA}
\affiliation{Department of Astronomy and Astrophysics, University of Chicago, 5640 South Ellis Avenue, Chicago, IL, 60637, USA}
\author{P.~M.~Chichura\,\orcidlink{0000-0002-5397-9035}}
\affiliation{Department of Physics, University of Chicago, 5640 South Ellis Avenue, Chicago, IL, 60637, USA}
\affiliation{Kavli Institute for Cosmological Physics, University of Chicago, 5640 South Ellis Avenue, Chicago, IL, 60637, USA}
\author{A.~Chokshi}
\affiliation{Department of Astronomy and Astrophysics, University of Chicago, 5640 South Ellis Avenue, Chicago, IL, 60637, USA}
\author{T.-L.~Chou\,\orcidlink{0000-0002-3091-8790}}
\affiliation{Department of Astronomy and Astrophysics, University of Chicago, 5640 South Ellis Avenue, Chicago, IL, 60637, USA}
\affiliation{Kavli Institute for Cosmological Physics, University of Chicago, 5640 South Ellis Avenue, Chicago, IL, 60637, USA}
\affiliation{National Taiwan University, No. 1, Sec. 4, Roosevelt Road, Taipei 106319, Taiwan}
\author{A.~Coerver\,\orcidlink{0000-0002-2707-1672}}
\affiliation{Department of Physics, University of California, Berkeley, CA, 94720, USA}
\author{T.~M.~Crawford\,\orcidlink{0000-0001-9000-5013}}
\affiliation{Department of Astronomy and Astrophysics, University of Chicago, 5640 South Ellis Avenue, Chicago, IL, 60637, USA}
\affiliation{Kavli Institute for Cosmological Physics, University of Chicago, 5640 South Ellis Avenue, Chicago, IL, 60637, USA}
\author{C.~Daley\,\orcidlink{0000-0002-3760-2086}}
\affiliation{Universit\'e Paris-Saclay, Universit\'e Paris Cit\'e, CEA, CNRS, AIM, 91191, Gif-sur-Yvette, France}
\affiliation{Department of Astronomy, University of Illinois Urbana-Champaign, 1002 West Green Street, Urbana, IL, 61801, USA}
\author{T.~de~Haan\,\orcidlink{0000-0001-5105-9473}}
\affiliation{High Energy Accelerator Research Organization (KEK), Tsukuba, Ibaraki 305-0801, Japan}
\author{K.~R.~Dibert}
\affiliation{Department of Astronomy and Astrophysics, University of Chicago, 5640 South Ellis Avenue, Chicago, IL, 60637, USA}
\affiliation{Kavli Institute for Cosmological Physics, University of Chicago, 5640 South Ellis Avenue, Chicago, IL, 60637, USA}
\author{M.~A.~Dobbs}
\affiliation{Department of Physics and McGill Space Institute, McGill University, 3600 Rue University, Montreal, Quebec H3A 2T8, Canada}
\affiliation{Canadian Institute for Advanced Research, CIFAR Program in Gravity and the Extreme Universe, Toronto, ON, M5G 1Z8, Canada}
\author{M.~Doohan}
\affiliation{School of Physics, University of Melbourne, Parkville, VIC 3010, Australia}
\author{D.~Dutcher\,\orcidlink{0000-0002-9962-2058}}
\affiliation{Joseph Henry Laboratories of Physics, Jadwin Hall, Princeton University, Princeton, NJ 08544, USA}
\author{C.~Feng}
\affiliation{Department of Astronomy, University of Science and Technology of China, Hefei 230026, China}
\affiliation{School of Astronomy and Space Science, University of Science and Technology of China, Hefei 230026}
\affiliation{Department of Physics, University of Illinois Urbana-Champaign, 1110 West Green Street, Urbana, IL, 61801, USA}
\author{K.~R.~Ferguson\,\orcidlink{0000-0002-4928-8813}}
\affiliation{Department of Physics and Astronomy, University of California, Los Angeles, CA, 90095, USA}
\affiliation{Department of Physics and Astronomy, Michigan State University, East Lansing, MI 48824, USA}
\author{N.~C.~Ferree\,\orcidlink{0000-0002-7130-7099}}
\affiliation{California Institute of Technology, 1200 East California Boulevard., Pasadena, CA, 91125, USA}
\affiliation{Kavli Institute for Particle Astrophysics and Cosmology, Stanford University, 452 Lomita Mall, Stanford, CA, 94305, USA}
\affiliation{Department of Physics, Stanford University, 382 Via Pueblo Mall, Stanford, CA, 94305, USA}
\author{K.~Fichman}
\affiliation{Department of Physics, University of Chicago, 5640 South Ellis Avenue, Chicago, IL, 60637, USA}
\affiliation{Kavli Institute for Cosmological Physics, University of Chicago, 5640 South Ellis Avenue, Chicago, IL, 60637, USA}
\author{A.~Foster\,\orcidlink{0000-0002-7145-1824}}
\affiliation{Joseph Henry Laboratories of Physics, Jadwin Hall, Princeton University, Princeton, NJ 08544, USA}
\author{S.~Galli}
\affiliation{Sorbonne Universit\'e, CNRS, UMR 7095, Institut d'Astrophysique de Paris, 98 bis bd Arago, 75014 Paris, France}
\author{A.~E.~Gambrel}
\affiliation{Kavli Institute for Cosmological Physics, University of Chicago, 5640 South Ellis Avenue, Chicago, IL, 60637, USA}
\author{A.~K.~Gao}
\affiliation{Department of Physics, University of Illinois Urbana-Champaign, 1110 West Green Street, Urbana, IL, 61801, USA}
\author{F.~Ge}
\affiliation{California Institute of Technology, 1200 East California Boulevard., Pasadena, CA, 91125, USA}
\affiliation{Kavli Institute for Particle Astrophysics and Cosmology, Stanford University, 452 Lomita Mall, Stanford, CA, 94305, USA}
\affiliation{Department of Physics, Stanford University, 382 Via Pueblo Mall, Stanford, CA, 94305, USA}
\affiliation{Department of Physics \& Astronomy, University of California, One Shields Avenue, Davis, CA 95616, USA}
\author{F.~Guidi\,\orcidlink{0000-0001-7593-3962}}
\affiliation{Department of Physics \& Astronomy, University of California, One Shields Avenue, Davis, CA 95616, USA}
\affiliation{Sorbonne Universit\'e, CNRS, UMR 7095, Institut d'Astrophysique de Paris, 98 bis bd Arago, 75014 Paris, France}
\author{S.~Guns}
\affiliation{Department of Physics, University of California, Berkeley, CA, 94720, USA}
\author{N.~W.~Halverson}
\affiliation{CASA, Department of Astrophysical and Planetary Sciences, University of Colorado, Boulder, CO, 80309, USA }
\affiliation{Department of Physics, University of Colorado, Boulder, CO, 80309, USA}
\author{E.~Hivon\,\orcidlink{0000-0003-1880-2733}}
\affiliation{Sorbonne Universit\'e, CNRS, UMR 7095, Institut d'Astrophysique de Paris, 98 bis bd Arago, 75014 Paris, France}
\author{G.~P.~Holder\,\orcidlink{0000-0002-0463-6394}}
\affiliation{Department of Physics, University of Illinois Urbana-Champaign, 1110 West Green Street, Urbana, IL, 61801, USA}
\author{W.~L.~Holzapfel}
\affiliation{Department of Physics, University of California, Berkeley, CA, 94720, USA}
\author{J.~C.~Hood}
\affiliation{Kavli Institute for Cosmological Physics, University of Chicago, 5640 South Ellis Avenue, Chicago, IL, 60637, USA}
\author{A.~Hryciuk}
\affiliation{Department of Physics, University of Chicago, 5640 South Ellis Avenue, Chicago, IL, 60637, USA}
\affiliation{Kavli Institute for Cosmological Physics, University of Chicago, 5640 South Ellis Avenue, Chicago, IL, 60637, USA}
\author{N.~Huang\,\orcidlink{0000-0003-3595-0359}}
\affiliation{Department of Physics, University of California, Berkeley, CA, 94720, USA}
\author{T.~Jhaveri}
\affiliation{Department of Astronomy and Astrophysics, University of Chicago, 5640 South Ellis Avenue, Chicago, IL, 60637, USA}
\affiliation{Kavli Institute for Cosmological Physics, University of Chicago, 5640 South Ellis Avenue, Chicago, IL, 60637, USA}
\author{F.~K\'eruzor\'e}
\affiliation{High-Energy Physics Division, Argonne National Laboratory, 9700 South Cass Avenue, Lemont, IL, 60439, USA}
\author{A.~R.~Khalife\,\orcidlink{0000-0002-8388-4950}}
\affiliation{Sorbonne Universit\'e, CNRS, UMR 7095, Institut d'Astrophysique de Paris, 98 bis bd Arago, 75014 Paris, France}
\author{L.~Knox}
\affiliation{Department of Physics \& Astronomy, University of California, One Shields Avenue, Davis, CA 95616, USA}
\author{K.~Kornoelje}
\affiliation{Department of Astronomy and Astrophysics, University of Chicago, 5640 South Ellis Avenue, Chicago, IL, 60637, USA}
\affiliation{Kavli Institute for Cosmological Physics, University of Chicago, 5640 South Ellis Avenue, Chicago, IL, 60637, USA}
\affiliation{High-Energy Physics Division, Argonne National Laboratory, 9700 South Cass Avenue, Lemont, IL, 60439, USA}
\author{C.-L.~Kuo}
\affiliation{Kavli Institute for Particle Astrophysics and Cosmology, Stanford University, 452 Lomita Mall, Stanford, CA, 94305, USA}
\affiliation{Department of Physics, Stanford University, 382 Via Pueblo Mall, Stanford, CA, 94305, USA}
\affiliation{SLAC National Accelerator Laboratory, 2575 Sand Hill Road, Menlo Park, CA, 94025, USA}
\author{K.~Levy}
\affiliation{School of Physics, University of Melbourne, Parkville, VIC 3010, Australia}
\author{Y.~Li\,\orcidlink{0000-0002-4820-1122}}
\affiliation{Kavli Institute for Cosmological Physics, University of Chicago, 5640 South Ellis Avenue, Chicago, IL, 60637, USA}
\author{A.~E.~Lowitz\,\orcidlink{0000-0002-4747-4276}}
\affiliation{Kavli Institute for Cosmological Physics, University of Chicago, 5640 South Ellis Avenue, Chicago, IL, 60637, USA}
\author{C.~Lu}
\affiliation{Department of Physics, University of Illinois Urbana-Champaign, 1110 West Green Street, Urbana, IL, 61801, USA}
\author{G.~P.~Lynch\,\orcidlink{0009-0004-3143-1708}}
\affiliation{Department of Physics \& Astronomy, University of California, One Shields Avenue, Davis, CA 95616, USA}
\author{T.~J.~Maccarone\,\orcidlink{0000-0003-0976-4755}}
\affiliation{Department of Physics \& Astronomy, Box 41051, Texas Tech University, Lubbock TX 79409-1051, USA}
\author{A.~S.~Maniyar\,\orcidlink{0000-0002-4617-9320}}
\affiliation{Kavli Institute for Particle Astrophysics and Cosmology, Stanford University, 452 Lomita Mall, Stanford, CA, 94305, USA}
\affiliation{Department of Physics, Stanford University, 382 Via Pueblo Mall, Stanford, CA, 94305, USA}
\affiliation{SLAC National Accelerator Laboratory, 2575 Sand Hill Road, Menlo Park, CA, 94025, USA}
\author{E.~S.~Martsen}
\affiliation{Department of Astronomy and Astrophysics, University of Chicago, 5640 South Ellis Avenue, Chicago, IL, 60637, USA}
\affiliation{Kavli Institute for Cosmological Physics, University of Chicago, 5640 South Ellis Avenue, Chicago, IL, 60637, USA}
\author{F.~Menanteau}
\affiliation{Department of Astronomy, University of Illinois Urbana-Champaign, 1002 West Green Street, Urbana, IL, 61801, USA}
\affiliation{Center for AstroPhysical Surveys, National Center for Supercomputing Applications, Urbana, IL, 61801, USA}
\author{M.~Millea\,\orcidlink{0000-0001-7317-0551}}
\affiliation{Department of Physics, University of California, Berkeley, CA, 94720, USA}
\author{J.~Montgomery}
\affiliation{Department of Physics and McGill Space Institute, McGill University, 3600 Rue University, Montreal, Quebec H3A 2T8, Canada}
\author{T.~Natoli}
\affiliation{Kavli Institute for Cosmological Physics, University of Chicago, 5640 South Ellis Avenue, Chicago, IL, 60637, USA}
\author{A.~Ouellette\,\orcidlink{0000-0003-0170-5638}}
\affiliation{Department of Physics, University of Illinois Urbana-Champaign, 1110 West Green Street, Urbana, IL, 61801, USA}
\author{Z.~Pan\,\orcidlink{0000-0002-6164-9861}}
\affiliation{High-Energy Physics Division, Argonne National Laboratory, 9700 South Cass Avenue, Lemont, IL, 60439, USA}
\affiliation{Kavli Institute for Cosmological Physics, University of Chicago, 5640 South Ellis Avenue, Chicago, IL, 60637, USA}
\affiliation{Department of Physics, University of Chicago, 5640 South Ellis Avenue, Chicago, IL, 60637, USA}
\author{P.~Paschos}
\affiliation{Enrico Fermi Institute, University of Chicago, 5640 South Ellis Avenue, Chicago, IL, 60637, USA}
\author{K.~A.~Phadke\,\orcidlink{0000-0001-7946-557X}}
\affiliation{Department of Astronomy, University of Illinois Urbana-Champaign, 1002 West Green Street, Urbana, IL, 61801, USA}
\affiliation{Center for AstroPhysical Surveys, National Center for Supercomputing Applications, Urbana, IL, 61801, USA}
\affiliation{NSF-Simons AI Institute for the Sky (SkAI), 172 E. Chestnut St., Chicago, IL 60611, USA}
\author{K.~Prabhu}
\affiliation{Department of Physics \& Astronomy, University of California, One Shields Avenue, Davis, CA 95616, USA}
\author{W.~Quan\,\orcidlink{0009-0002-2589-5501}}
\affiliation{High-Energy Physics Division, Argonne National Laboratory, 9700 South Cass Avenue, Lemont, IL, 60439, USA}
\affiliation{Department of Physics, University of Chicago, 5640 South Ellis Avenue, Chicago, IL, 60637, USA}
\affiliation{Kavli Institute for Cosmological Physics, University of Chicago, 5640 South Ellis Avenue, Chicago, IL, 60637, USA}
\author{S.~Raghunathan\,\orcidlink{0000-0003-1405-378X}}
\affiliation{Department of Physics \& Astronomy, University of California, One Shields Avenue, Davis, CA 95616, USA}
\affiliation{Center for AstroPhysical Surveys, National Center for Supercomputing Applications, Urbana, IL, 61801, USA}
\author{M.~Rahimi}
\affiliation{School of Physics, University of Melbourne, Parkville, VIC 3010, Australia}
\author{A.~Rahlin\,\orcidlink{0000-0003-3953-1776}}
\affiliation{Department of Astronomy and Astrophysics, University of Chicago, 5640 South Ellis Avenue, Chicago, IL, 60637, USA}
\affiliation{Kavli Institute for Cosmological Physics, University of Chicago, 5640 South Ellis Avenue, Chicago, IL, 60637, USA}
\author{C.~L.~Reichardt\,\orcidlink{0000-0003-2226-9169}}
\affiliation{School of Physics, University of Melbourne, Parkville, VIC 3010, Australia}
\author{M.~Rouble}
\affiliation{Department of Physics and McGill Space Institute, McGill University, 3600 Rue University, Montreal, Quebec H3A 2T8, Canada}
\author{J.~E.~Ruhl}
\affiliation{Department of Physics, Case Western Reserve University, Cleveland, OH, 44106, USA}
\author{A.~C.~Silva~Oliveira\,\orcidlink{0000-0001-5755-5865}}
\affiliation{California Institute of Technology, 1200 East California Boulevard., Pasadena, CA, 91125, USA}
\affiliation{Kavli Institute for Particle Astrophysics and Cosmology, Stanford University, 452 Lomita Mall, Stanford, CA, 94305, USA}
\affiliation{Department of Physics, Stanford University, 382 Via Pueblo Mall, Stanford, CA, 94305, USA}
\author{A.~Simpson}
\affiliation{Department of Astronomy and Astrophysics, University of Chicago, 5640 South Ellis Avenue, Chicago, IL, 60637, USA}
\affiliation{Kavli Institute for Cosmological Physics, University of Chicago, 5640 South Ellis Avenue, Chicago, IL, 60637, USA}
\author{J.~A.~Sobrin\,\orcidlink{0000-0001-6155-5315}}
\affiliation{Department of Physics, Villanova University, 800 E Lancaster Ave, Villanova, PA 19085, USA}
\affiliation{Fermi National Accelerator Laboratory, MS209, P.O. Box 500, Batavia, IL, 60510, USA}
\author{A.~A.~Stark}
\affiliation{Center for Astrophysics \textbar{} Harvard \& Smithsonian, 60 Garden Street, Cambridge, MA, 02138, USA}
\author{J.~Stephen}
\affiliation{Enrico Fermi Institute, University of Chicago, 5640 South Ellis Avenue, Chicago, IL, 60637, USA}
\author{C.~Tandoi\,\orcidlink{0000-0002-2077-6004}}
\affiliation{Department of Astronomy, University of Illinois Urbana-Champaign, 1002 West Green Street, Urbana, IL, 61801, USA}
\author{C.~Trendafilova}
\affiliation{CERCA/ISO, Department of Physics, Case Western Reserve University, Cleveland, OH 44106, USA}
\affiliation{Center for AstroPhysical Surveys, National Center for Supercomputing Applications, Urbana, IL, 61801, USA}
\author{J.~D.~Vieira\,\orcidlink{0000-0001-7192-3871}}
\affiliation{Department of Astronomy, University of Illinois Urbana-Champaign, 1002 West Green Street, Urbana, IL, 61801, USA}
\affiliation{Department of Physics, University of Illinois Urbana-Champaign, 1110 West Green Street, Urbana, IL, 61801, USA}
\affiliation{Center for AstroPhysical Surveys, National Center for Supercomputing Applications, Urbana, IL, 61801, USA}
\author{A.~G.~Vieregg\,\orcidlink{0000-0002-4528-9886}}
\affiliation{Kavli Institute for Cosmological Physics, University of Chicago, 5640 South Ellis Avenue, Chicago, IL, 60637, USA}
\affiliation{Department of Astronomy and Astrophysics, University of Chicago, 5640 South Ellis Avenue, Chicago, IL, 60637, USA}
\affiliation{Enrico Fermi Institute, University of Chicago, 5640 South Ellis Avenue, Chicago, IL, 60637, USA}
\affiliation{Department of Physics, University of Chicago, 5640 South Ellis Avenue, Chicago, IL, 60637, USA}
\author{A.~Vitrier\,\orcidlink{0009-0009-3168-092X}}
\affiliation{Sorbonne Universit\'e, CNRS, UMR 7095, Institut d'Astrophysique de Paris, 98 bis bd Arago, 75014 Paris, France}
\author{Y.~Wan}
\affiliation{Department of Astronomy, University of Illinois Urbana-Champaign, 1002 West Green Street, Urbana, IL, 61801, USA}
\affiliation{Center for AstroPhysical Surveys, National Center for Supercomputing Applications, Urbana, IL, 61801, USA}
\author{N.~Whitehorn\,\orcidlink{0000-0002-3157-0407}}
\affiliation{Department of Physics and Astronomy, Michigan State University, East Lansing, MI 48824, USA}
\author{M.~R.~Young}
\affiliation{Fermi National Accelerator Laboratory, MS209, P.O. Box 500, Batavia, IL, 60510, USA}
\affiliation{Kavli Institute for Cosmological Physics, University of Chicago, 5640 South Ellis Avenue, Chicago, IL, 60637, USA}
\author{J.~A.~Zebrowski}
\affiliation{Kavli Institute for Cosmological Physics, University of Chicago, 5640 South Ellis Avenue, Chicago, IL, 60637, USA}
\affiliation{Department of Astronomy and Astrophysics, University of Chicago, 5640 South Ellis Avenue, Chicago, IL, 60637, USA}
\affiliation{Fermi National Accelerator Laboratory, MS209, P.O. Box 500, Batavia, IL, 60510, USA}
\collaboration{SPT-3G Collaboration}
\noaffiliation